\documentclass[aps,twocolumn,superscriptaddress,floatfix,showpacs]{revtex4} 
\usepackage{amsmath,euscript,theorem,graphicx,rotating}
\usepackage[latin1]{inputenc}
\usepackage[T1]{fontenc}
\usepackage{txfonts}
\usepackage{url}
\usepackage{array,dcolumn}
\usepackage{xr} 
\usepackage{color}
\usepackage{dcolumn}
\usepackage{bm}
\usepackage{longtable}
\usepackage{xspace}
\usepackage{booktabs}
\usepackage{version}

\bibstyle{achemso}

\usepackage{array}
\newcolumntype{L}{>{$}l<{$}}
\newcolumntype{R}{>{$}r<{$}}
\newcolumntype{C}{>{$}c<{$}}

\newcommand{\rr}{{\mathbf r}}

\newcommand{\RR}{{\mathbf R}}

\newcommand{\etal}{\emph{et al.}\xspace}

\newcommand{\order}[1]{\ensuremath{\mathcal{O}(#1)}\xspace}
\newcommand{\neighbourhood}[1]{\ensuremath{\mathcal{N}_{#1}}\xspace}
\newcommand{\kJpermol}{\ensuremath{\rm{kJ\,mol}^{-1}}\xspace}

\newcommand{\rms}[0]{r.m.s.\xspace}

\newcommand{\Eelstpen}[0]{\ensuremath{E^{(1)}_{\rm pen}}\xspace}

\newcommand{\Eelst}[0]{\ensuremath{E^{(1)}_{\rm elst}}\xspace}
\newcommand{\EelstMP}[0]{\ensuremath{V^{(1)}_{\rm elst}[{\rm DM}]}\xspace}
\newcommand{\Eexch}[0]{\ensuremath{E^{(1)}_{\rm exch}}\xspace}
\newcommand{\EINDreg}[1]{\ensuremath{E^{(#1)}_{\rm IND}({\rm Reg})}\xspace}
\newcommand{\EIND}[1]{\ensuremath{E^{(#1)}_{\rm IND}}\xspace}
\newcommand{\EpolMP}[1]{\ensuremath{V^{(#1)}_{\rm pol}[{\rm DM}]}\xspace}

\newcommand{\Eindpol}[1]{\ensuremath{E^{(#1)}_{\rm ind,pol}}\xspace}

\newcommand{\Eexind}[1]{\ensuremath{E^{(#1)}_{\rm ind,exch}}\xspace}

\newcommand{\EdispMP}[1]{\ensuremath{V^{(#1)}_{\rm disp}[{\rm DM}]}\xspace}
\newcommand{\EDISP}[1]{\ensuremath{E^{(#1)}_{\rm DISP}}\xspace}
\newcommand{\Edisppol}[1]{\ensuremath{E^{(#1)}_{\rm disp,pol}}\xspace}
\newcommand{\Eexdisp}[1]{\ensuremath{E^{(#1)}_{\rm disp,exch}}\xspace}

\newcommand{\Eint}[1]{\ensuremath{E_{\rm int}^{(#1)}}\xspace}
\newcommand{\Esr}[1]{\ensuremath{E_{\rm sr}^{(#1)}}\xspace}
\newcommand{\ESR}[0]{\ensuremath{E_{\rm sr}}\xspace}

\newcommand{\deltaHF}[0]{\ensuremath{\delta^{\rm HF}_{\rm int}}\xspace}

\newcommand{\VINT}[0]{\ensuremath{V_{\text{int}}}\xspace}
\newcommand{\Vint}[1]{\ensuremath{V_{\text{int}}[#1]}\xspace}
\newcommand{\Ves}[1]{\ensuremath{V_{\text{elst}}[#1]}\xspace}

\newcommand{\Vpol}[1]{\ensuremath{V_{\text{pol}}[#1]}\xspace}
\newcommand{\Vdisp}[1]{\ensuremath{V_{\text{disp}}[#1]}\xspace}
\newcommand{\VSR}[1]{\ensuremath{V^{#1}_{\rm sr}}\xspace}
\newcommand{\Vsr}[2]{\ensuremath{V^{#1}_{\rm sr}[#2]}\xspace}
\newcommand{\Cn}[1]{\ensuremath{C_{#1}}\xspace}

\newcommand{\Cniso}[1]{\ensuremath{C_{#1}{\rm (iso)}}\xspace}

\newcommand{\CnisoOPT}[1]{\ensuremath{C_{#1}{\rm (iso,opt)}}\xspace}
\newcommand{\tCnisoOPT}[1]{\ensuremath{\tilde{C}_{#1}{\rm (iso,opt)}}\xspace}

\newcommand{\ECT}[1]{\ensuremath{E_{\rm CT}^{(#1)}}\xspace}
\newcommand{\EPOL}[1]{\ensuremath{E_{\rm POL}^{(#1)}}\xspace}

\newcommand{\BETAelst}[1]{\ensuremath{\beta_{\rm elst}^{#1}}\xspace}
\newcommand{\betapol}[0]{\ensuremath{\beta_{\rm pol}}\xspace}
\newcommand{\BETApol}[1]{\ensuremath{\beta_{\rm pol}^{#1}}\xspace}
\newcommand{\betadisp}[0]{\ensuremath{\beta_{\rm disp}}\xspace}
\newcommand{\BETAdisp}[1]{\ensuremath{\beta_{\rm disp}^{#1}}\xspace}

\newcommand{\CamCASP}{{\sc CamCASP}\xspace}
\newcommand{\ORIENT}{{\sc Orient}\xspace}

\newcommand{\Orient}{\ORIENT}
\newcommand{\SAPT}{{\sc Sapt2008}\xspace}

\newcommand{\DALTON}{{\sc DALTON} 2.0\xspace}
\newcommand{\Dalton}{\DALTON}
\newcommand{\NWChem}{{\sc NWChem} 6.x\xspace}
\newcommand{\Gamess}{{\sc GAMESS(US)} \xspace}
\newcommand{\Gaussian}{{\sc Gaussian03} \xspace}

\newcommand{\abinitio}{{\em ab initio}\xspace}
\newcommand{\Abinitio}{{\em Ab initio}\xspace}

\newcommand{\w}[1]{\ensuremath{w^{#1}}\xspace}

\newcommand{\wL}[1]{\ensuremath{w_{\rm L}^{#1}}\xspace}
\newcommand{\wT}[1]{\ensuremath{\tilde{w}^{#1}}\xspace}

\mathchardef\lt="313C \mathchardef\gt="313E
\mathcode`<="4268 \mathcode`>="5269
\newcolumntype{d}[1]{D{.}{.}{#1}}
\newcolumntype{.}{D{.}{.}{-1}}
\newcolumntype{,}{D{,}{,}{2}}

\newcommand{\JCP}[0]{J. Chem. Phys.\ }
\newcommand{\JPCA}[0]{J. Phys. Chem. A\ }
\newcommand{\JPCB}[0]{J. Phys. Chem. B\ }

\newcommand{\JCTC}[0]{J. Chem. Theory Comput.\ }
\newcommand{\IJQC}[0]{Int. J. Quantum Chem.\ }
\newcommand{\CPL}[0]{Chem. Phys. Lett.\ }
\newcommand{\TCA}[0]{Theor. Chim. Acta\ }

\newcommand{\PRB}[0]{Phys. Rev. B\ }

\newcommand{\PRL}[0]{Phys. Rev. Lett.\ }

\newcommand{\MolP}[0]{Mol. Phys.\ }

\newcommand{\PCCP}[0]{Phys. Chem. Chem. Phys.\ }

\newcommand{\JACS}[0]{J. Am. Chem. Soc.\ }

\newcommand{\IRPC}[0]{Int. Revs. Phys. Chem.\ }

\newcommand{\ProcRoyalSocA}[0]{Proc. R. Soc. A\ }

\newcommand{\SurfSciLett}[0]{Surf. Sci. Lett.\ }

\newcommand{\paperB}{Part II\xspace}
\begin{document}

\title[CamCASP]
{
  {\em Ab initio} atom--atom potentials using \CamCASP : 
  Theory and application to multipole models for the pyridine dimer.
}

\author{Alston J. Misquitta}
\affiliation{School of Physics and Astronomy, Queen Mary, University of London,
London E1 4NS, U.K.}
\email{a.j.misquitta@qmul.ac.uk}
\author{Anthony J. Stone}
\affiliation{University Chemical Laboratory, Lensfield Road,
Cambridge, CB2 1EW, U.K.}

\date{\today}

\begin{abstract}
    Creating accurate, analytic atom--atom potentials for small organic molecules
    from first principles can be a time-consuming and computationally intensive
    task, particularly if we also require them to include
    explicit polarization terms, which are essential in many systems.
    In this first part of a two-part investigation, we describe how the
    \CamCASP suite of programs can be used to generate such potentials using
    some of the most accurate electronic structure methods currently applicable.
    In particular, we derive the long-range terms from
    monomer properties, and determine the 
    short-range anisotropy parameters by a novel and robust method
    based on the iterated stockholder atom approach. 
    We use the techniques described here to develop distributed multipole models for the 
    electrostatic, polarization and dispersion interactions in
    the pyridine dimer.
    In the second part of this work we will apply these methods to 
    develop a series of many-body potentials for the pyridine system.
\end{abstract}

\maketitle

\section{Introduction}
\label{sec:introduction}

Electronic structure methods for the interaction energy have come a long way
since the mid-nineties, when the water dimer represented the largest
system for which accurate, \abinitio intermolecular interaction
energies could be calculated.
We can now calculate interaction energies for small organic molecules
like pyridine and benzene in hours on a single processor 
\cite{PodeszwaBS06a,HesselmannJS05,HesselmannJS06a}, and medium sized molecules
like cyclotrimethylene trinitramine (RDX) \cite{PodeszwaBRS07},
base pairs \cite{HesselmannJS06}, and tetramers of amino acids \cite{FiethenJHS08}.
Part of the reason for this is the increase in our computational
resources, but more important are the new developments in electronic structure
methods. For the field of intermolecular interactions, the 
development of symmetry-adapted perturbation theory based on 
density-functional theory, or SAPT(DFT), has done much to improve
both the accuracy and the range of applicability of theoretical 
methods.
\cite{MisquittaS02,MisquittaJS03,MisquittaS05,MisquittaPJS05b,PodeszwaBS06a,
      HesselmannJ02a,HesselmannJ02b,HesselmannJ03a,HesselmannJS05}

However, such calculations cannot be used on the fly 
in most molecular simulations, as the computational cost is too high,
and we need to represent the interaction energy by an analytic potential.
%
%
Such potentials are commonly expressed in terms of the many-body
expansion, where
the interaction energy of a cluster of interacting molecules
is partitioned into two-body contributions plus corrections
arising from triplets, quartets and larger clusters of molecules.
That is,
\begin{equation}
  V_{ABC\dots}      = \sum_{X \lt Y} V_{XY}
                    +  \sum_{X \lt Y \lt Z} \Delta V_{XYZ}
                    + \cdots,
\end{equation}
where $V_{XY}$ is the interaction energy between
molecules $X$ and $Y$ in the absence of all other molecules, but in
the geometry found in the complete system, while
$\Delta V_{XYZ}$ is the three-body 
correction, defined as 
\begin{equation*}
\Delta V_{XYZ} = V_{XYZ} - V_{XY} - V_{XZ} - V_{YZ}
\label{eq:mbexp}
\end{equation*}
and $V_{XYZ}$ is the energy of the $XYZ$ cluster in the
absence of all other molecules, but in the geometry found in the
complete system.
Four-body, five-body and other many-body corrections are defined in a
similar manner.

The success of this expansion depends on its rapid convergence.
In any molecular system with distinct interacting units, the two-body terms
will dominate, but the many-body terms can contribute as much as 30\% of the
interaction energy for clusters of polar molecules \cite{HodgesSX97,MasBS03a,MasBS03b},
and can be essential for getting the structure and properties correct.
For example, three and four-body effects have been shown to be responsible
for the tetrahedral structure of liquid water \cite{BukowskiSGvanderA06}.
The many-body polarization energy has also been shown to be an 
important discriminator in the relative lattice energies of 
molecular crystals when the structures differ considerably in their
hydrogen-bonding motifs \cite{WelchKMSP08}.

A three-body implementation of SAPT(DFT)
does exist \cite{PodeszwaS07}, but the computational cost makes
on-the-fly methods even more impractical, and although
three-body non-additive interactions make up the bulk of the many-body 
non-additivity in systems like water, non-additive effects beyond this
level cannot be neglected \cite{BukowskiSGvanderA06}.
If the constituent bodies in a cluster are small enough, it would be possible
to use an electronic structure method like SAPT(DFT) or CCSD(T)
(coupled-cluster singles and doubles with non-iterated triples) for the two
and three-body terms in the many-body expansion, and an appropriate
approximation for the other terms. 
But more generally this approach would
make formidable computational demands, and it is necessary to use
analytic intermolecular potentials in many applications.

Analytic intermolecular potentials have been in use for many decades.
(See ref.~\citenum{Stone:book:13} for a review.)
In the past, most have been `pair potentials',
including only two-body terms.
In any molecular system with distinct interacting units, the two-body terms
will dominate, but the many-body terms can be essential for getting
the structure and properties correct. 
The effects of many-body terms have often been
included in an approximate `average' manner through adjustment of the empirical parameters.
This is done in empirical potentials for water, which typically feature
an enhanced dipole moment to mimic the increased average dipole of the water molecule in the 
condensed phase.
While such pair potentials are still widely used, it is increasingly
recognised that it is necessary to take account of the many-body
effects explicitly, particularly to account for the effects of
electrostatic polarization \cite{JiangJT07,WelchKMSP08,IllingworthD09},
but also to account for many-body dispersion effects \cite{BukowskiSJJSKWR99,BukowskiS01,KimSVCL06},
and, as we shall see, to account for intermolecular charge delocalisation,
or charge transfer (CT).

Potentials with this level of complexity, accuracy and detail cannot be 
obtained empirically. Instead we must turn to theoretical methods.
\Abinitio-derived potentials are by no means new, and 
indeed there are a number of accurate potentials in the
published literature (see for example
refs.~\citenum{KoronaWBJS97,BukowskiSC99,MisquittaBS00,VissersHJWvanderA05}).
These potentials have typically been obtained for small dimers, but recently examples
involving medium sized systems have become available
\cite{PodeszwaBS06b,PodeszwaBRS07,MisquittaWSP08,TottonMK10a}.
There are a few common ideas used in the creation of these and other \abinitio potentials. 
The first is that they are all based on a distributed model; that is, the interaction
energy between molecules is represented as the sum of contributions
between pairs of atoms. Secondly, most are not
polarizable, so many-body polarization terms are missing (though
polarization may be
included at the two-body level). Thirdly, in all cases, long-range parameters have been 
derived from the unperturbed molecules, which can dramatically simplify the number of free 
parameters in the fit. Finally, the short-range parameters are usually then fitted to a
set of \abinitio interaction energies calculated using a suitable electronic structure 
method.

The above procedure works reasonably well, but it has a number of deficiencies. 
First and foremost is the usual lack of many-body 
polarization effects. Second, there is much uncertainty 
associated with fitting the short-range exponential terms in a system of medium 
sized molecules. These uncertainties are largely related to sampling: we are usually
not sure that we have enough data to define the terms in the potential. This is 
particularly troublesome for the larger systems, which not only have 
a larger number of free parameters to fit, but which also incur considerable computational
expense to calculate the \abinitio interaction energies needed for the fit.
Additionally, the short-range terms are usually exponential in form,
and it is very difficult to fit a sum of exponentials while 
also requiring that the fit parameters remain physically sensible
and transferable. Some of these difficulties can be partially alleviated by
iterating the process and adding additional data at important configurations
\cite{PodeszwaBS06b}, but on the whole this approach is unsatisfactory and
tedious, and an alternative is needed.

The alternative we describe in this paper is to compute directly most of the potential
parameters, including those associated with the short-range part of the potential, 
and keep the fitting to a minimum. In many ways this is not a new strategy; indeed, 
a similar technique has been implemented by Schmidt and co-workers
\cite{YuMcDS11,McDanielS13,SchmidtYMcD15}, who have used many of the techniques
we will describe in this paper to develop a family of transferable potentials
with a strong physical basis. However, so far these have been isotropic potentials
of moderate accuracy, with a strong focus on ease of creation and transferability.
As we will demonstrate here, we bring a new level of fidelity, accuracy and reliability to the
procedure, using the many tools we have developed in
recent years and have implemented in the \CamCASP \cite{CamCASP}
program. We begin this paper with a description of the overall strategy,
then describe some of the algorithms we have implemented in the \CamCASP
suite of programs to implement the strategy. In particular,
partitioning the electron density using the iterated stockholder atom
procedure is very effective in overcoming the difficulties in fitting
the short-range potential. In the second paper we apply these methods
to the pyridine dimer and discuss the resulting potentials.

\section{The problem and definitions}
\label{sec:problem}

The goal is to find an analytic potential \VINT that accurately models
the two-body SAPT(DFT) interaction energy
\begin{align}
    \Eint{1-\infty} &= \Eelst + \Eexch + \EIND{2} + \EDISP{2} + \deltaHF.
\end{align}
(We will use $E$ throughout to denote the computed energy terms and $V$ to denote their
analytic representations.)
Here \Eelst and \Eexch are the first-order electrostatic and exchange-repulsion
energies, $\EIND{2} = \Eindpol{2} + \Eexind{2}$ is the total second-order
induction energy, $\EDISP{2} = \Edisppol{2} + \Eexdisp{2}$ is the total 
dispersion energy \cite{Jansen15private}, and $\deltaHF$ is the estimate of effects
of third and higher order, primarily induction \cite{JeziorskaJC87,MoszynskiHJ96}.
The broad strategy we have adopted to determine \VINT has been described in some 
detail in a review article \cite{StoneM07}. While many of the details have changed,
the essence of the method remains as described there, so only a high-level description
will be provided here. 

First of all, we represent the potential \VINT as 
\begin{align}
    \VINT &= \sum_{a \in A} \sum_{b \in B} \Vint{ab}(r_{ab},\Omega_{ab}),
\end{align}
where, $a$ and $b$ label sites (usually taken to be atomic sites) in the interacting
molecules $A$ and $B$, $r_{ab}$ is the inter-site separation, 
$\Omega_{ab}$ is a suitable set of angular coordinates that describes the relative 
orientation of the local axis systems on these sites 
(see ch.\ 12 in ref.~\citenum{Stone:book:13}), and \Vint{ab} is the site--site
potential defined as 
\begin{equation}
    \Vint{ab} = \Vsr{}{ab} + \Ves{ab} + \Vdisp{ab} + \Vpol{ab}.
    \label{eq:Vtot}
\end{equation}
The terms in $\Vint{ab}$ model the corresponding terms in
$\Eint{1-\infty}$. 
\Vsr{}{ab} is the short-range term, which mainly describes the
exchange--repulsion energy, but also includes some other short-range
effects, discussed in \S\ref{sec:shortrange}:
\begin{equation}
    \Vsr{}{ab} = G \exp{[-\alpha_{ab}(\Omega_{ab})(r_{ab} - \rho_{ab}(\Omega_{ab}))]},
    \label{eq:Vtot_Vsr}
\end{equation}
where $\rho_{ab}(\Omega_{ab})$ is the shape function for this pair of
sites, which depends on their relative orientation $\Omega_{ab}$, and
$\alpha_{ab}$ is the 
hardness parameter which may also be a function of orientation. $G$ is
a constant energy which we will take to be $10^{-3}$ hartree.
$\Ves{ab}$ is the expanded electrostatic energy:
\begin{equation}
    \Ves{ab} = \Ves{ab}\bigl( r_{ab},\Omega_{ab}, Q^{a}_{t}, Q^{b}_{u},
        \BETAelst{ab} \bigr);
    \label{eq:Vtot_Ves}
\end{equation}
$Q^{a}_{t}$ is the multipole moment of rank
$t$ for site $a$, where, using the compact notation of ref.~\citenum{Stone:book:13}, 
$t=00,10,11c,11s,\cdots$, and $\BETAelst{ab}$ is
a damping parameter. 
The dispersion energy $\Vdisp{ab}$ depends on the anisotropic
dispersion coefficients  $C_n^{ab}(\Omega_{ab})$ for the pair of
sites, and on a damping function $f_n$ that we will take to be
the Tang--Toennies \cite{TangT92} incomplete gamma functions of order $n+1$: 
\begin{equation}
    \Vdisp{ab} =  - \sum_{n = 6}^{12} f_n\Bigl(\BETAdisp{ab} r_{ab}\Bigr) 
        C_n^{ab}(\Omega_{ab})r_{ab}^{-n}
    \label{eq:Vtot_Vdisp}
\end{equation}
The final term $\Vpol{ab}$ is the polarization energy, which is the long-range
part of the induction energy \cite{Misquitta13a}. $\Vpol{ab}$  depends on
the multipole moments and the polarizabilities $\alpha^{a}_{tu}$,
which are indexed by pairs of multipole components $tu$ 
(for details see refs.\citenum{MisquittaS08a,Stone:book:13}):
\begin{equation}
    \Vpol{ab} = \Vpol{ab}\bigl(Q^{a}_{t}, Q^{b}_{u}, \alpha^{a}_{tu},
                \alpha^{b}_{tu},\BETApol{ab} \bigr).
    \label{eq:Vtot_Vpol}
\end{equation}
 
There are a few points to note about the particular form of the potential \Vpol{ab}.
Although formally written in the form of a two-body potential, many-body polarization
effects are included through the classical polarization expansion \cite{Stone:book:13}.
Also, we will normally define the multipole moments and polarizabilities to
include \emph{intra}molecular many-body effects implicitly, that is, we use the multipoles and 
polarizabilities of atoms-in-a-molecule, localized appropriately.
To this form of the potential we could add a three-body dispersion model, but this is 
not addressed in this paper.

\section{Strategy}
\label{sec:strategy}

There are many parameters in such a potential and our goal is to \emph{compute}
as many of these parameters as possible, and keep the fitting of the remainder
to a minimum. Additionally, we will adopt a hierarchical approach to the fitting
process that helps to guarantee confidence in the parameter values.
There are three main parts to the process, and these involve the following:
\begin{itemize}
  \item \emph{Long-range terms}: The electrostatic, polarization and dispersion
    interaction energy components possess expansions in powers of $1/R$, where
    $R$ is the centre-of-mass separation (for small systems) or, more
    generally, the inter-site distance in a distributed expansion. 
    Multipole moments are functions of the unperturbed molecular densities and
    may be derived using a variety of methods, the most common being the
    distributed multipole analysis (DMA) technique \cite{StoneA85,Stone05b}.
    But, using a basis-space algorithm of the iterated stockholder atom (ISA) 
    procedure\cite{LillestolenW09} termed the BS-ISA algorithm \cite{MisquittaSF14},
    we have demonstrated that the ISA/BS-ISA-based distribution yields a more rapidly
    convergent multipole expansion with properties that make it ideal for modelling.
    The distributed polarizabilities and dispersion coefficients are obtained
    using the Williams--Stone--Misquitta (WSM) technique 
    \cite{MisquittaS06,MisquittaS08a,MisquittaSP08,MisquittaS08b}. 
    With this approach we may consider the long range parameters in the potential
    \VINT as fixed, though, we may optionally tune them if appropriate. 
  \item \emph{Damping}: All three multipole expansions need to be damped at short range,
    when overlap effects become appreciable and the $1/R$ terms start to exhibit
    mathematical divergences. Damping will not be applied to the electrostatic 
    expansion as it is not usually needed, but it can be applied if necessary \cite{Stone11}.
    It is crucial to damp the polarization and dispersion expansions as the 
    mathematical divergence of these expansions is usually manifest at 
    accessible separations, and must be controlled if
    sensible expansions are needed. For the dispersion expansion we use
    a single damping coefficient based on the vertical ionization potentials
    $I_A$ and $I_B$ (measured in atomic units) of the interacting molecules
    \cite{MisquittaS08b}:
    \begin{align}
        \BETAdisp{ab} \equiv \BETAdisp{AB} = \sqrt{2 I_A} + \sqrt{2 I_B}.
        \label{eq:disp-damping}
    \end{align}
    This single-parameter damping is almost certainly not ideal, and we should
    rather use damping parameters that depend on the atomic types, and optionally, 
    on their relative orientation. We will propose such a more elaborate, but 
    still non-empirical model in a forthcoming paper \cite{VanVleetMSS15}.

    The damping of the polarization expansion is less straightforward and will
    be discussed in detail below.
  \item \emph{Short-range energies}: If the damped multipole (DM) expanded energies are removed
    from the interaction energy $\Eint{1-\infty}$, we obtain the remainder which is 
    the short-range energy:
    \begin{align}
        \Esr{1-\infty} &= \Eexch + ( \Eelst - \EelstMP ) \nonumber \\
                       &\quad + ( \EIND{2} + \deltaHF - \EpolMP{2-\infty} ) \notag\\
                       &\quad + ( \EDISP{2} - \EdispMP{2} ) \nonumber \\
        &= \Esr{1} + \Esr{2-\infty}.
                      \label{eq:Esr}
    \end{align}
    Here we have partitioned the short-range energy into a first-order contribution
    $\Esr{1}$ which will be dominant, and the second- to infinite-order contribution
    $\Esr{2-\infty}$ which will be primarily the infinite-order charge-transfer energy.
    In the above expression, $\EelstMP$ and $\EdispMP{2}$ are the multipole expanded forms
    of the electrostatic and dispersion energies, and $\EpolMP{2-\infty}$ is the 
    infinite-order (iterated) multipole-expanded polarization energy.
    In principle, the various contributions to $\Esr{1-\infty}$ are not expected to
    depend on dimer geometry in the same way and they should be modelled separately.
    However, we have previously showed that the dominant contributions to \Esr{1}
    ---the first-order exchange and penetration energies---are
    proportional to each other,\cite{MisquittaSF14} and here we will show that the charge-transfer
    contribution is also nearly proportional, so we shall 
    model all parts of $\Esr{1-\infty}$ together as a single sum of exponential terms:
    \begin{align}
        \VSR{}  &= \sum_{a \in A} \sum_{b \in B}  \Vsr{}{ab}
                      \label{eq:Vsr}
        \end{align}
        where each $\Vsr{}{ab}$ has the form of eq.~\eqref{eq:Vtot_Vsr}.

  \item \emph{Sampling dimer configuration space}:
     In order to ensure a balanced fit, it is important to ensure that we sample 
     the six dimensional dimer configuration space adequately. For such a 
     high dimensional space the sampling needs to be (quasi) random, and in 
     earlier work \cite{MisquittaWSP08,StoneM07,TottonMK10a} we have described
     how this can be done using a quasi random Sobol sequence and Shoemake's 
     algorithm \cite{Shoemake92} (see the supplementary information for a brief description
     of this algorithm). This algorithm has been implemented in the
     \CamCASP program and ensures that we cover orientation space randomly,
     but uniformly. Unless otherwise indicated, dimer configurations will be
     obtained using this algorithm. 

  \item \emph{Fitting the short-range terms: first-order energies}: 
    A direct fit to the terms in \VSR{} usually leads to unphysical parameters and
    therefore should be avoided. Additionally, it is difficult to
    sample the high-dimensional configuration space densely enough to 
    define the shape anisotropy of the interacting sites.
    This is particularly true for the larger molecular systems, for which the computational
    cost of calculating the second to infinite order SAPT(DFT) interaction energies
    can be appreciable, thus precluding the possibility of adequate sampling.
    One possibility in this case is to reduce the complexity of
    \VSR{} by, for example,
    keeping only isotropic terms in the expansions for the hardness parameter and the
    shape functions, but this may not be appropriate when high accuracies are needed.

    In previous work \cite{MisquittaWSP08} we addressed this problem using the density-overlap
    model \cite{KitaNI76,KimKL81,NobeliPW98} to partition the first-order short-range energies, $\Esr{1}$,
    into contributions from pairs of atoms. This partitioning allows us to fit the terms
    for each pair of sites $ab$ and obtain a first guess at \Vsr{(1)}{ab}, while avoiding fitting the
    sum of exponential terms directly. 
    In \S\ref{sec:dens-overlap-model} we provide more detail on how this is done, 
    and show how the parameters in eq.~\eqref{eq:Vsr} can be determined with
    a high degree of confidence if we use a density
    partitioning method based on the ISA method. 
    As we shall see, this procedure effectively eliminates the basis-set
    limitations seen in our earlier attempts. Moreover, this step uses the 
    first-order energies only, and these energies are not only computationally inexpensive,
    but may be calculated using a monomer basis set, so a dense coverage of 
    configuration space may be used to determine good initial guesses for the parameters in \VSR{(1)}.
    In this manner, atomic shape functions may be determined easily and reliably. 

  \item \emph{Constrained relaxation}: 
    At various stages in the fitting process we will relax a fit with constraints applied.
    The idea here is to obtain a good guess for the parameters in the fit in a manner 
    that ensures that they are well-defined. Subsequently, these parameters may be relaxed
    while pinning them to the predetermined values. Consider a fitting function $g(p_0, p_1,\cdots,p_n)$,
    where $p_i$ are the free parameters in the fit. If our initial guess for these are
    $p^0_i$, then in a constrained relaxation we would optimize the function
    \begin{align}
        G(p_0, p_1,\cdots,p_n) &= g(p_0, p_1,\cdots,p_n)\notag\\
        &\qquad+ \sum_{i=0}^{n} c_i (p_i - p^0_i)^2,
        \label{eq:constrained-opt}
    \end{align}
    where $c_i$ are suitable constraint strength parameters that should be
    associated with our confidence in 
    the initial parameter guesses $p^0_i$. In a Bayesian sense, the ${p^0_i}$ are our prior 
    values and the ${c_i}$ will be related to the prior distribution. As data is included,
    the parameters ${p_i}$ may deviate from their initial values. In this manner, a fit 
    may be performed with very little data and we ensure that no parameter attains an unphysical 
    value.

  \item \emph{Relaxing \VSR{(1)} to $\Esr{1}$}:
    Having obtained the first guess for \VSR{(1)}, we may now perform a constrained
    relaxation of the parameters in \VSR{(1)} to fit $\Esr{1}$ better. 
    Symmetry constraints to the shape-function parameters may also
    be imposed at this stage.

  \item \emph{Relaxing \VSR{(1)} to include higher-order energies}:
    The parameters in \VSR{(1)} may now be further relaxed to account for the 
    higher order short-range energies, $\Esr{2-\infty}$, thereby obtaining the 
    full short-range potential \VSR{}. The higher-order short-range energies
    will normally
    be evaluated on a much sparser set of points, so the constraints
    used in this relaxation step usually need to be fairly tight, and
    the anisotropy terms should probably be kept fixed at this stage unless
    enough data can be made available.

  \item \emph{Overall relaxation and iterations}: 
    The relaxation steps may be repeated as additional data is added. This is a common
    strategy, but here we do the relaxation with fairly tight constraints.
    Additional dimer energies are best calculated at special configurations on the 
    potential energy surface. These would include stable minima and regions of
    configuration space near minima. A suitably \emph{converged} fit is one which 
    is stable with respect to the inclusion of additional data. 
\end{itemize}

Some of these steps have already
been used to create accurate \abinitio potentials \cite{MisquittaWSP08,TottonMK10a}, and
indeed, some of these ideas have been used and developed by other research groups 
(see for example, Refs. \citenum{NobeliPW98,PodeszwaBS06b,McDanielYS12}).
What is unique to this work is the manner in which these steps have been combined with 
advanced density-partitioning methods, distribution techniques and a hierarchical calculation
of intermolecular interaction energies, so as to obtain intermolecular interaction potentials
easily and reliably and with high accuracy. We describe most of these
steps in detail below, and will elaborate further on those related to the short-range
potential in \paperB.

\section{Numerical details}
\label{sec:numerical}

The geometry of the pyridine molecule was optimized using the \Gaussian 
program\cite{Gaussian03} using the PBE0 functional \cite{AdamoB99a}
and the cc-pVTZ Dunning basis set \cite{Dunning89}. The $C_{2v}$
point group symmetry was imposed during the optimization. 

\subsection{Comments on the kinds of basis sets}
We use several kinds of basis sets to calculate the various data needed
for the intermolecular potential of pyridine. The SAPT(DFT) interaction
energies require diffuse monomer basis sets augmented with mid-bond
basis functions to converge the dispersion energy, and additionally
basis functions located on the partner monomer -- the so called far-bond 
functions --- to converge the charge-transfer component of the 
induction energy. The resulting basis is referred to as the MC+ basis type
\cite{WilliamsMSJ95}.
The $\deltaHF$ term requires a calculation of the super-molecular interaction
energy at the Hartree--Fock level, and therefore needs to be calculated
using a dimer-centered basis. In both cases the density-fitting needed
for the SAPT and SAPT(DFT) energies is done in a dimer-centered auxiliary basis,
possibly augmented with a suitable mid-bond set.
For high accuracies the Cartesian form of the auxiliary basis is used.

We compute the large set of first-order energies in a
monomer-centered basis and subsequently rotate all quantities to the
required dimer orientation. However, for accurate first-order interaction 
energies, the auxiliary basis used in these calculations must still
be the dimer-centered type. Additionally, in this case we use the spherical
form of the basis functions as the \CamCASP programme is, as yet, unable
to rotate objects calculated using Cartesian functions.

Monomer properties are normally calculated in a monomer-centered basis that
is taken to be the monomer part of the basis set used for the SAPT(DFT)
energies. However this is not optimal as the additional off-atomic basis
functions used in the MC+ basis form have the effect of increasing
the size of the equivalent monomer-centered basis set. 
Consequently, it is advantageous to calculate the monomer properties in a
larger, more diffuse monomer basis as this would better match the 
multipole expanded energies with those from the non-expanded 
SAPT(DFT) calculations.

\subsection{Basis set details}
The distributed molecular properties were calculated using asymptotically
corrected PBE0 (PBE0/AC) with the d-aug-cc-pVTZ Dunning basis \cite{WoonD94}.
The density-functional calculation was performed using a modified version of
the \Dalton program \cite{DALTON2} with modifications made using a patch
provided as part of the \SAPT suite of programs \cite{SAPT2008}.
The asymptotic correction was performed using the Fermi--Amaldi long-range
form of the exchange potential with the Tozer--Handy splicing function
\cite{TozerH98} and a vertical ionization potential of 0.3488 a.u., calculated
using a $\Delta$-DFT procedure with the PBE0 functional and an aug-cc-pVTZ
basis set.
The \CamCASP program \cite{CamCASP} was used to evaluate the distributed 
multipole moments using both DMA and ISA algorithms, and distributed
static and frequency-dependent polarizabilities and dispersion coefficients
using the WSM algorithm \cite{MisquittaS06,MisquittaS08a,MisquittaSP08,MisquittaS08b}.
For the ISA calculation the auxiliary basis was constructed from the RI-MP2 aug-cc-pVQZ 
fitting basis \cite{WeigendHPA98,WeigendKH02} with $s$-functions replaced with those
from ISA-set2 supplied with the \CamCASP program \cite{MisquittaSF14}. 

Interaction energy calculations using SAPT(DFT) were performed using the 
\CamCASP program with molecular orbitals and eigenvalues calculated
with the \Dalton program using the PBE0/AC functional described above.
Second-order SAPT(DFT) interaction energy calculations were performed using
the Sadlej-pVTZ basis in the MC+ format (monomer basis plus mid-bond
functions) with a $3s2p1d$ mid-bond set 
\cite{BukowskiSJJSKWR99} placed on a site determined using a dispersion-weighted
algorithm \cite{AkinojoBS03}. The DC+ form of the RI-MP2 aug-cc-pVTZ auxiliary
basis \cite{WeigendHPA98,WeigendKH02} with Cartesian GTOs was used for density-fitting
with a $3s3p3d2f1g$ fitting mid-bond set with exponents 
$s$:~(1.1061,0.5017,0.2342), $p$:~(0.94,0.5,0.25), $d$:~(0.9,0.6,0.3),
$f$:~(0.7,0.4), $g$:~(0.65).
The hybrid ALDA+CHF kernel was used in all SAPT(DFT) calculations. Kernel integrals
were calculated initially within the \Dalton program, but subsequently they were
computed internally in \CamCASP with the ALDA part of the kernel constructed from
Slater exchange components and PW91 correlation kernel \cite{PerdewW92}.
The $\deltaHF$ correction was evaluated using the DC+ form of the Sadlej pVTZ basis 
with a corresponding DC+ auxiliary basis set formed from the RI-MP2 aug-cc-pVTZ
fitting basis and $3s3p3d2f1g$ fitting mid-bond set.

Additionally, first-order SAPT(DFT) interaction energies used
in the initial stage of the fit were calculated using a  monomer-centered
(MC) Sadlej-pVTZ basis \cite{Sadlej88} and a dimer-centered (DC) RI-MP2
aug-cc-pVTZ fitting basis \cite{WeigendHPA98,WeigendKH02}. The density-functional
calculations on the monomers were performed once using the PBE0/AC functional,
and the molecular orbitals were suitably rotated within \CamCASP for subsequent
first-order interaction energy calculations. For this purpose, due to current
requirements within \CamCASP, the spherical form of the Gaussian-type orbitals
(GTOs) was used for the auxiliary basis set.

\subsection{Data sets}
\label{sec:datasets}

In \S\ref{sec:strategy} we have described how intermolecular 
interaction potentials may be developed in multiple stages, with more accurate,
but less extensive data sets used in each successive stage. 
The pyridine dimer potentials we develop in this paper and in \paperB 
have been developed using three data sets:
\begin{itemize}
    \item \emph{Dataset(0)}: 
        First-order energies calculated on a set of 3515 pseudo-random dimer
        geometries obtained using Shoemake's algorithm as described above.
        This data set was used in the first stage of the fitting 
        process to obtain the initial short-range parameters using 
        the distributed density-overlap model. 
    \item \emph{Dataset(1)}: Infinite-order SAPT(DFT) interaction energies
        calculated on a set of 500 pseudo-random dimers also obtained using Shoemake's 
        algorithm. This data set was used in refining the dispersion model, in
        fitting the charge-transfer contribution to the interaction energy,
        and, in the final stage, to tune the total interaction energy models.
    \item \emph{Dataset(2)}: Infinite-order SAPT(DFT) interaction energies
        calculated on a set of 257 dimers obtained as special points (minima)
        from early versions of the pyridine potential development. These dimers
        are significantly lower in energy than those from Dataset(1).
        This data set served two purposes: Firstly, as it contained dimer geometries
        significantly different from those found in Dataset(1), it provided us with an 
        independent means of assessing the quality of the fits.
        Secondly, in the final stage, this data set was used to tune the total
        interaction energy models.
\end{itemize}

\section{Long-range methods}
\label{sec:longrange}

One of the fundamental advantages of intermolecular perturbation theories
like SAPT and\break SAPT(DFT) over supermolecular methods is that the energy 
components from perturbation theory have well-defined multipole
expansions \cite{JeziorskiS02}. Therefore the long-range form
of these energies can be derived from molecular properties such
as the multipole moments and static and frequency-dependent density-response
functions. This has the advantage that the asymptotic part of the 
potential energy surface is obtained directly, that is, without fitting.
Additionally, the long-range potential parameters are fully consistent with
the short-range energies from the perturbation theory.

In the \CamCASP suite of programs, we have implemented a number of
algorithms for calculating the distributed forms of the long-range
expansions of the electrostatic, polarization (induction) and dispersion
energies. The algorithms permit a considerable degree of freedom in the
model, so models may be more or less complex as the application requires.
The long-range terms in the model can be derived directly from monomer
properties, but there is a conflict between accuracy and computational
efficiency. 
We will aim to model most of the contributions to the interaction
energy separately, using several versions ranging from accurate but
computationally expensive to less accurate but cheaper.
For example, electrostatic models may be constructed using multipole models
from rank 0 (charges only) up to rank 4; or mixed rank models may also
be considered, with high ranking multipoles included only on some sites. 
This allows a considerable degree of flexibility in constructing the total
interaction energy model.
For this approach to work, we will need to ensure that each part of the model
is sufficiently accurate, with accuracy measured in a meaningful manner. 
Typically, we will expect to reduce \rms errors against some SAPT(DFT) reference
to less than 1 \kJpermol, and preferably less than 0.5 \kJpermol.

\subsection{Electrostatic models}
\label{sec:electrostatics}

Distributed multipole analysis is a well established procedure for
obtaining accurate electrostatic models from an \abinitio
wavefunction. We use the revised version of the
procedure\cite{Stone05b} which reduces the dependence of the multipole
description 
on basis set, at the cost of longer computation times.
This procedure uses a scheme based on
real-space grids for the density contributions arising from the diffuse 
functions, while for the more compact functions in the basis the original 
scheme is used. In this work the parameter controlling the switch
between compact and diffuse functions is set at 4.0, so the method is
denoted DMA4.

Until recently, the DMA approach has been the standard for distributed moments,
but recently we have demonstrated \cite{MisquittaSF14} that the 
ISA-based distributed multipole analysis (ISA-DMA) forms a significantly better
basis for potential
development as it guarantees fast and systematic convergence with 
respect to the rank of the expansion and a well-defined basis limit to the 
multipole components, and yields penetration energies (calculated as the difference
from the non-expanded \Eelst) more strongly proportional to the
first-order exchange energy \Eexch.
The last aspect of the ISA-DMA is particularly useful in model building, since
the proportionality of the electrostatic penetration energy to the first-order
exchange-repulsion energy allows us to combine the two and model their sum
with a single function. 
For the purposes of this paper we will define the electrostatic penetration energy
as \cite{MisquittaSF14}
\begin{align}
    \Eelstpen = \Eelst - \EelstMP,
      \label{eq:E1pen}
\end{align}
where \EelstMP denotes the electrostatic energy calculated from the 
distributed multipole (DM) expansion evaluated at convergence, which we will
take to be the model with terms from ranks 0 (charge) to 4 (hexadecapole). 

\begin{figure}
    \begin{center}
        \includegraphics[scale=0.45]{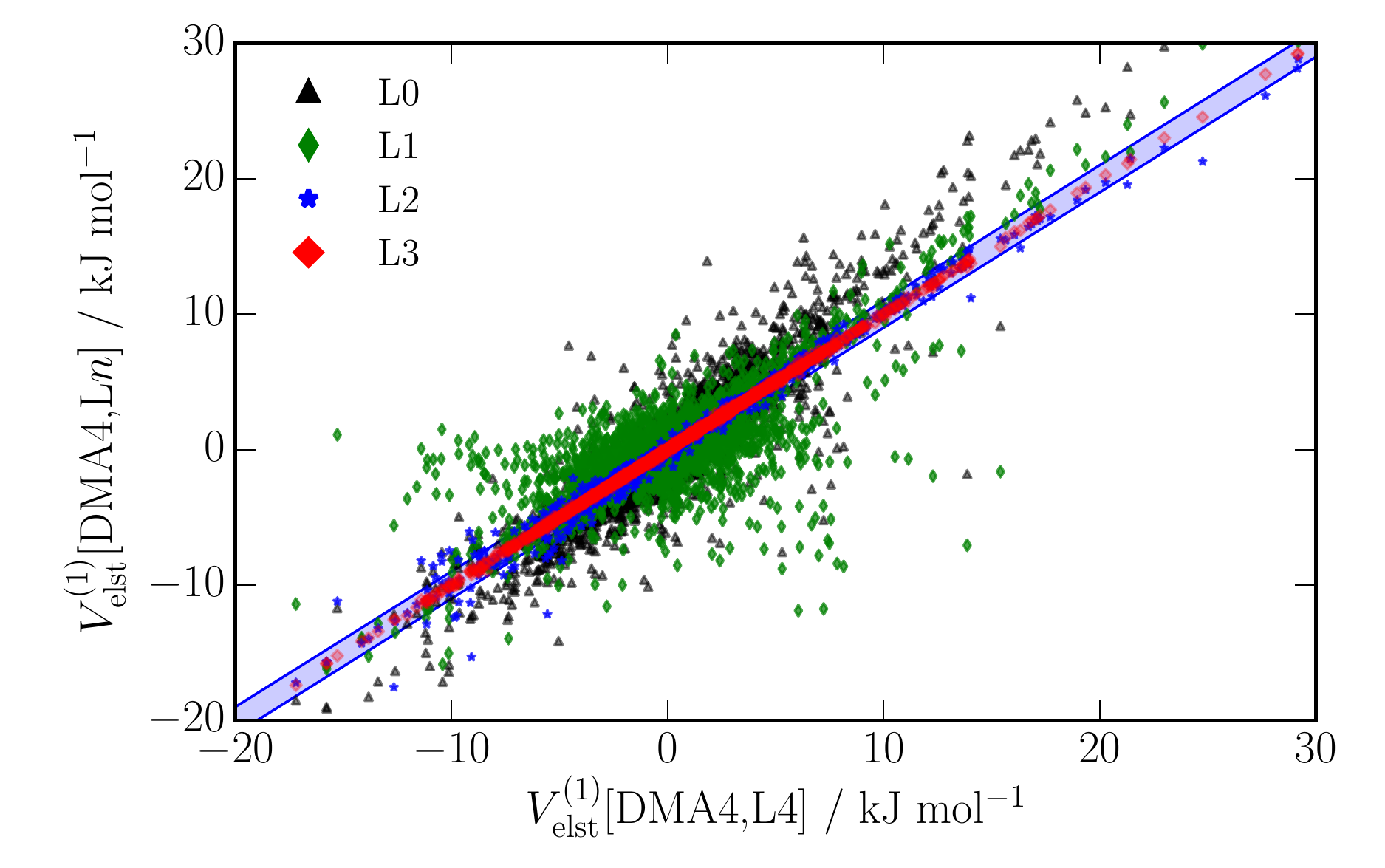}
        \includegraphics[scale=0.45]{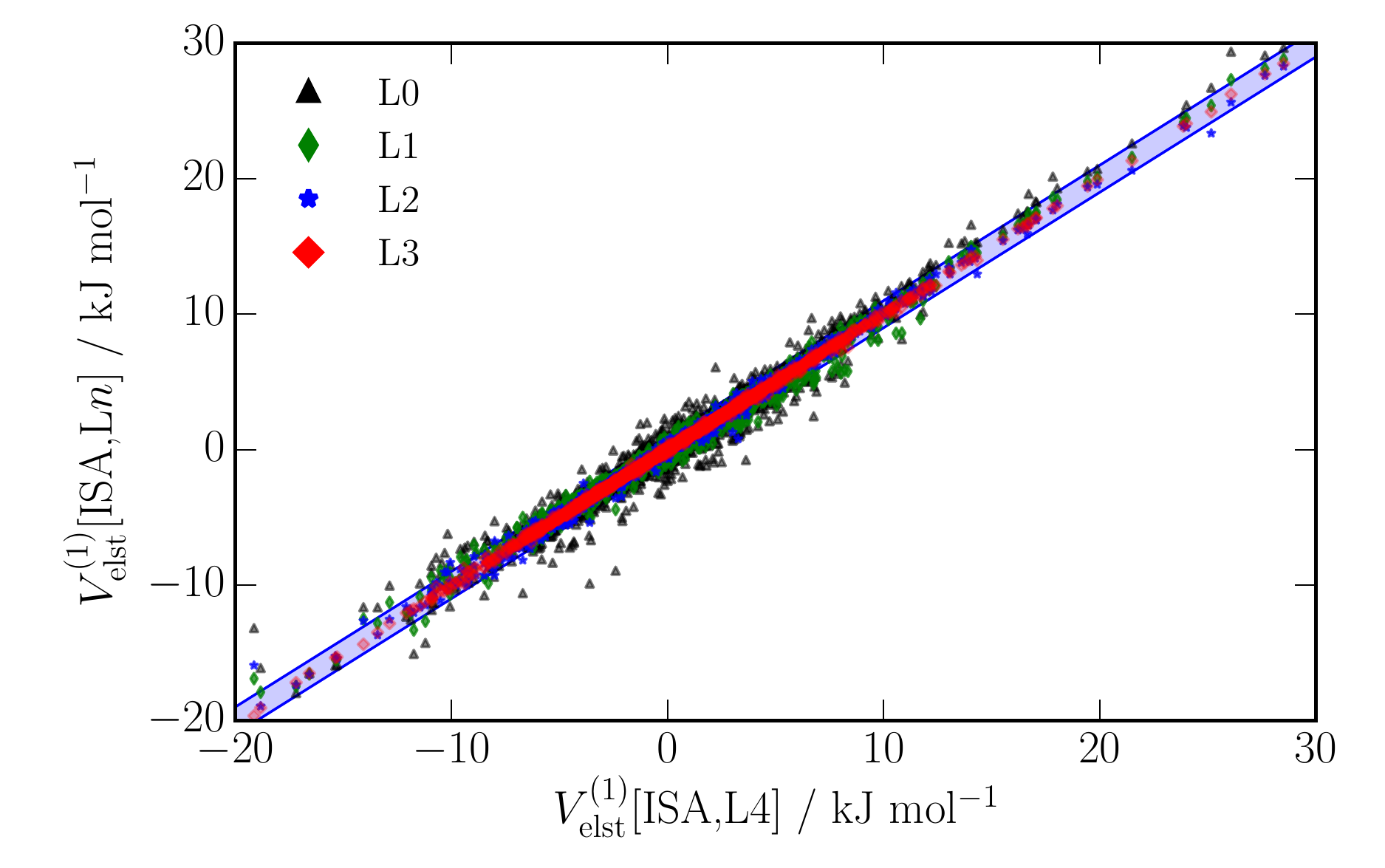}
    \end{center}
\caption{Scatter plot of model electrostatic energies from the DMA4 and ISA-DMA
    models at various ranks. The multipole expanded electrostatic energies \EelstMP
    for rank $n$ models, $n=0,1,2,3$, (i.e. including multipole
    moments only up to rank $n$) are plotted against the energies calculated
    with the rank~4 model (on the $x$-axis).
    No damping has been used. 
    The DMA4 results are in the top panel and the ISA-DMA (BS-ISA, $\zeta=0.1$) results
    are presented in the bottom panel.
    The blue bar represents the $\pm 1$ \kJpermol error range.
}
\label{fig:DMA4-ISA-elst-convergence}
\end{figure}

In fig.~7 of ref.~\citenum{MisquittaSF14} we demonstrated this aspect of the
ISA-DMA moments: in contrast to the DMA4 moments, the penetration energy
derived from the ISA-DMA model at rank 4 is indeed significantly more proportional
to \Eexch for the pyridine dimer. This alone makes the ISA-DMA model
more appropriate for this system---or indeed, any other,
as this proportionality seems to be generally true. 
Here we will look at the data presented in ref.~\citenum{MisquittaSF14} 
differently, to show more clearly how rapidly the DMA4 and ISA-DMA multipole
expansions converge with rank. 

For the construction of accurate electrostatic models, it is advisable
to include atom charges, dipoles and quadrupoles. The dipoles are
needed to describe features such as lone pairs, while quadrupoles are
needed to describe $\pi$-orbital features. Octopoles and hexadecapoles
can improve the description further but the improvement is not
generally worth the increased computational cost of the model.
However, for many applications, particularly for large molecules, due to 
program design limitations or more fundamentally, due to computational 
limitations, only charge models may be permissible. 
So the question arises: How do the multipole models behave when truncated
to lower orders in rank?
In Figure~\ref{fig:DMA4-ISA-elst-convergence} we have plotted \EelstMP calculated
with each of the two multipole models with truncated rank against the same with
all terms to rank 4 (deemed to be converged) included. We clearly see that while 
the rank 4 terms are not needed in the DMA4 model, any further truncation results
in unacceptably large errors and very little correlation is left between 
the converged results (terms to rank 4) and those with ranks limited to 0 (charges)
and 1 (charges and dipoles). 
In contrast, the ISA-DMA multipoles are distinctly better behaved upon truncation, with 
a strong correlation between all truncated models and the fully converged 
energies. This has some advantages: it may be possible to truncate the ISA-based 
distributed multipole model to much lower rank, perhaps even to rank 0, without
the need to re-parametrize the potential. 
We shall return to this issue below. 

We point out here that while the DMA4 multipole model is not directly
amenable to rank truncation, there is a way to perform a rank transformation
that generally does not result in significant errors.
This is done using by optimizing a distributed-multipole
description using the \textsc{Mulfit} program of Ferenczy
\emph{et al.}\cite{WinnFR97,FerenczyWR97}, in which the effects of
higher-rank multipoles on each atom are represented approximately by
multipoles of lower ranks on neighbouring atoms. In this way, a model
including multipoles up to quadrupole can incorporate some of the
effects of higher multipoles. This approach has recently been used
effectively to generate simple electrostatic models for a wide range
of polycyclic aromatic hydrocarbons occurring in the formation of
soot.\cite{TottonMK10a,TottonMK11} However the ISA-DMA treatment is
consistently better.

\subsection{Polarization and charge-transfer}
\label{sec:polarization}

In this paper we distinguish between the \emph{polarization} energy
and the \emph{induction} energy. In SAPT (or SAPT(DFT)), the polarization
energy and charge-transfer are combined in the induction energy.
We use regularised SAPT \cite{PatkowskiJS01a} to separate these two 
contributions \cite{Misquitta13a}, and by polarization energy
we mean that part of the induction energy that is not associated with
charge transfer.

The importance of polarizability in the interactions between polar and
polarizable molecules is now well recognized \cite{WelchKMSP08,SebetciB10a},
as is the inadequacy of the common approximation of
polarization effects by the use of enhanced static dipole moments.
In \CamCASP we use coupled Kohn--Sham perturbation theory to obtain an
accurate charge-density susceptibility, $\alpha(\rr,\rr')$, which
describes the change in charge density at $\rr$ in response to a
change in electrostatic potential at $\rr'$. 
Using a constrained density-fitting-based approach \cite{MisquittaS06},
the charge density susceptibility is partitioned
between atoms to obtain a distributed-polarizability model
$\alpha^{ab}_{tu}$ that gives the change in multipole $Q^b_u$ on atom
$b$ in response to a change in the electrostatic potential derivative $V^a_t$
at atom $a$. Here $u=00$ for the charge, $10=z$, $11c=x$ or $11s=y$ for
the dipole, $20$, $21c$, $21s$, $22c$ or $22s$ for the quadrupole
components, and so on; while $t=00$ for the electrostatic potential,
$10$, $11c$ or $11s$ for the components of the electrostatic field,
$20$ etc. for the field gradient, and so on. Note that the electric
field components are $E_x=E_{11c}=-V_{11c}$, $E_y=E_{11s}=-V_{11s}$
and $E_z=E_{10}=-V_{10}$.

This is a \emph{non-local} model of polarizability. That is, the
electric field at one atom of a molecule can induce a change in the
multipole moments on other atoms of the same molecule. This is an impractical and
unnecessarily complicated description that seems to be needed only for
special cases such as low-dimensional extended systems \cite{MisquittaSSA10}.
For most finite systems, the moments induced on
neighbouring atoms $b$ by a change in electric field on atom $a$ can
be represented by multipole expansions on atom $a$, giving a
\emph{local} polarizability description in which the effect of a
change in electric field at atom $a$ is described by changes in
multipole moments on that atom alone. This is a somewhat
over-simplified description of the procedure, and more detailed
accounts have been given by Stone \& Le Sueur\cite{LeSueurS94}, and by
Lillestolen \& Wheatley\cite{LillestolenW07}. The latter is a more
elaborate approach that deals rather better with the convergence
issues arising from induced moments on atoms distant from the one on
which the perturbation occurs. The local polarizability model is a
much more compact 
and useful description. In particular, the local picture removes
charge-flow effects, where a difference in potential between two atoms
induces a flow of charge between them. Such flows of charge still occur,
but they are described in terms of local dipole polarizabilities.
We point out here that the `self-repulsion plus local orthogonality'
(SRLO) distribution method \cite{RobS13a} can be
used to eliminate the charge-flow terms altogether (for most molecules).
This technique, which is a modification of the constrained density-fitting-based
distribution method \cite{MisquittaS06} is available in \CamCASP but
has not been used for the results of this paper. The SRLO polarizabilities
are non-local and will typically need localization to be usable by most
simulation programs. 

The resulting localized polarizability description can be refined by
the method of Williams \& Stone \cite{WilliamsS03} using the 
point-to-point responses: the
change in potential at each of an array of points around the molecule
in response to a point charge at any of the points. 
An important advantage of this method is that the final, 
refined polarization model can be chosen to suit the
problem---for example a simple isotropic dipole--dipole model, or an
elaborate model with anisotropic polarizabilities up to
quadrupole--quadrupole or higher. 
For a given choice of model, the refinement procedure ensures that we
obtain the highest accuracy (in an unbiased sense if sufficiently dense
grids of point-to-point responses are used) subject to the limitations
of the model. 
The combination of the SAPT(DFT) calculation of local (point-to-point) 
responses with this refinement procedure is referred to here as the
WSM method \cite{MisquittaS08a,MisquittaSP08}.

\begin{figure}
    \begin{center}
        \includegraphics[viewport=20 230 600 650,scale=0.7,clip]{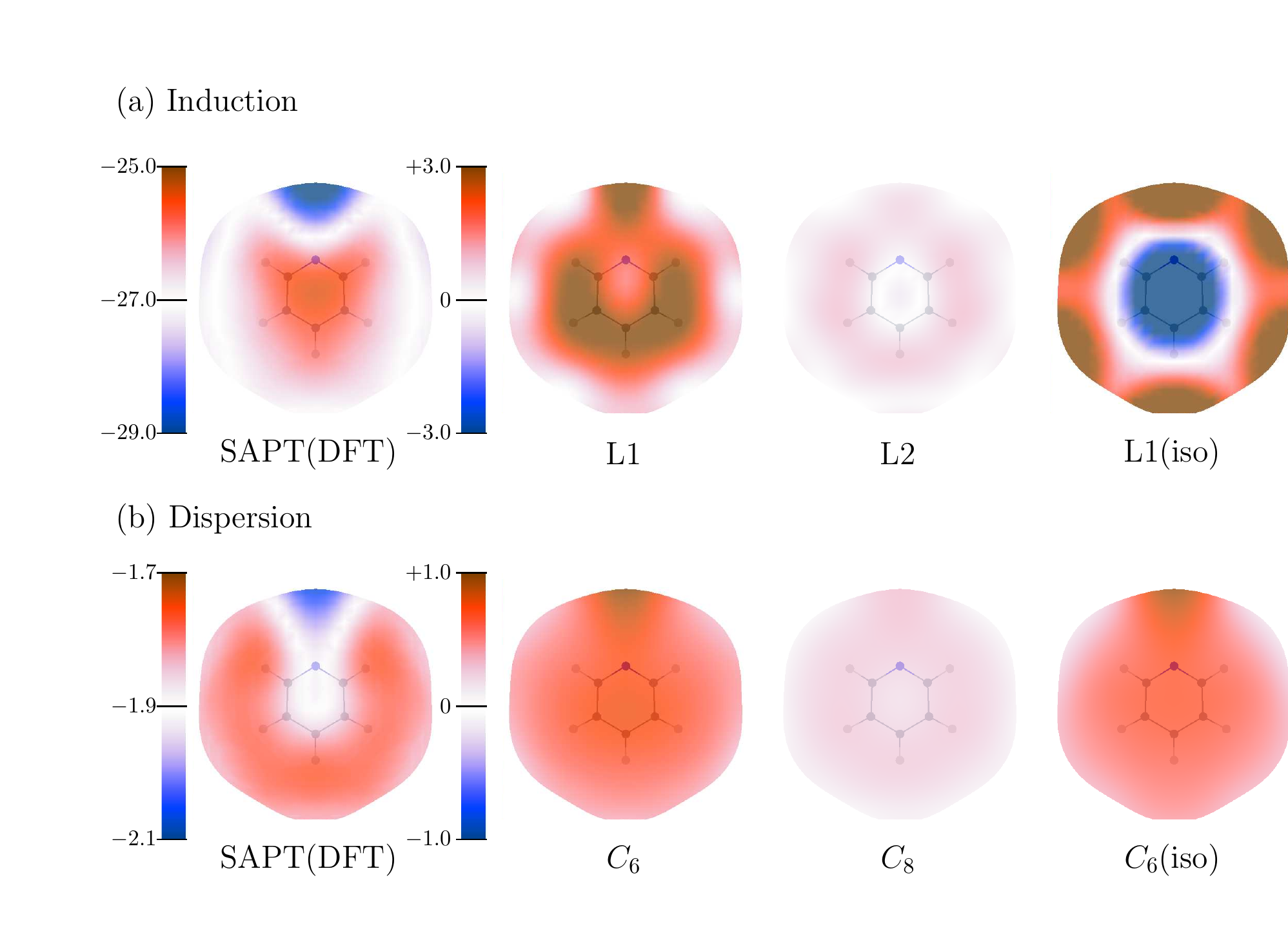}
    \end{center}
\caption{(a) Polarization and (b) dispersion energy maps and difference maps on
the 2$\times$vdW surface of pyridine.
Polarization energies have been calculated using a $+1e$ point-charge probe and
dispersion energies with a neon atom probe. Energies in \kJpermol.
}
\label{fig:maps}
\end{figure}

The quality of the WSM description can be judged by the accuracy of the
interaction energy of a point charge with the molecule. This
interaction comprises the classical electrostatic energy of
interaction of the point charge with the molecular charge
distribution, and the additional term, the polarization energy, that
arises from the relaxation of the molecular charge distribution in
response to the point charge. These components can be separated using
SAPT(DFT). The polarization energy of pyridine in the field of a
point charge is mapped in the left-hand picture of Figure~\ref{fig:maps}(a).
We construct a grid on the vdW$\times2$ surface of pyridine---that is, the surface
made up of spheres of twice the van der Waals radius around each
atom---and the polarization energy is calculated for a unit point charge at each
point of the grid in turn. The remaining three maps in
Figure~\ref{fig:maps}(a) show the error in the polarization energy for
three local polarizability descriptions: L1 uses dipole
polarizabilities on each atom, L2 includes dipole--quadrupole and
quadrupole--quadrupole polarizabilities, and
L1,iso uses isotropic dipole polarizabilities on each atom. It is clear
that the dipole-polarizability models are rather poor, and that an
accurate description needs to include quadrupole polarizabilities.

\subsubsection{Polarization damping}
\label{sec:pol-damping}

If the polarization interaction between molecules is calculated using
distributed multipoles for the electrostatic potential and distributed
polarizabilities for the polarization model, the effects of molecular
overlap are absent and damping is needed to avoid the so-called polarization
catastrophe which results in unphysical energies.
In our early work on this issue \cite{MisquittaS08a} we advocated damping the
classical polarization expansion to best match the total induction energies from
SAPT(DFT). 
Through numerical simulations of the condensed phase and the work of 
Sebetci and Beran \cite{SebetciB10a} we now know this to be incorrect, as it
leads to excessive many-body polarization energies.
The polarization damping must instead be determined by requiring that the classical
polarization model energies best match the true polarization energies from 
SAPT(DFT) \cite{Misquitta13a}. As noted above, perturbation theories like SAPT and SAPT(DFT) 
do not define a true polarization energy, but rather the 
induction energy, which is the sum of the polarization energy and the
charge-transfer energy. Recently one of us 
described how regularized SAPT(DFT) can be used to split the second-order
induction energy into the second-order polarization and charge-transfer
components \cite{Misquitta13a} which are defined as follows:
\begin{align}
    \EPOL{2} &= \EINDreg{2} \nonumber \\
    \ECT{2}  &= \EIND{2} - \EINDreg{2},
\end{align}
where \EINDreg{2} is the regularized second-order induction energy.
This definition leads to a well-defined basis limit for the second-order
polarization and charge-transfer energies \cite{Misquitta13a}. 
We determine the damping needed for the classical polarization expansion
by requiring that the non-iterated model energies best match \EPOL{2}. 
Once a suitable damping has been found, an estimate for the infinite-order
polarization energy \EPOL{2-\infty} is obtained by iterating the classical polarization
model to convergence.

\begin{figure}
    \begin{center}
        \includegraphics[scale=0.45]{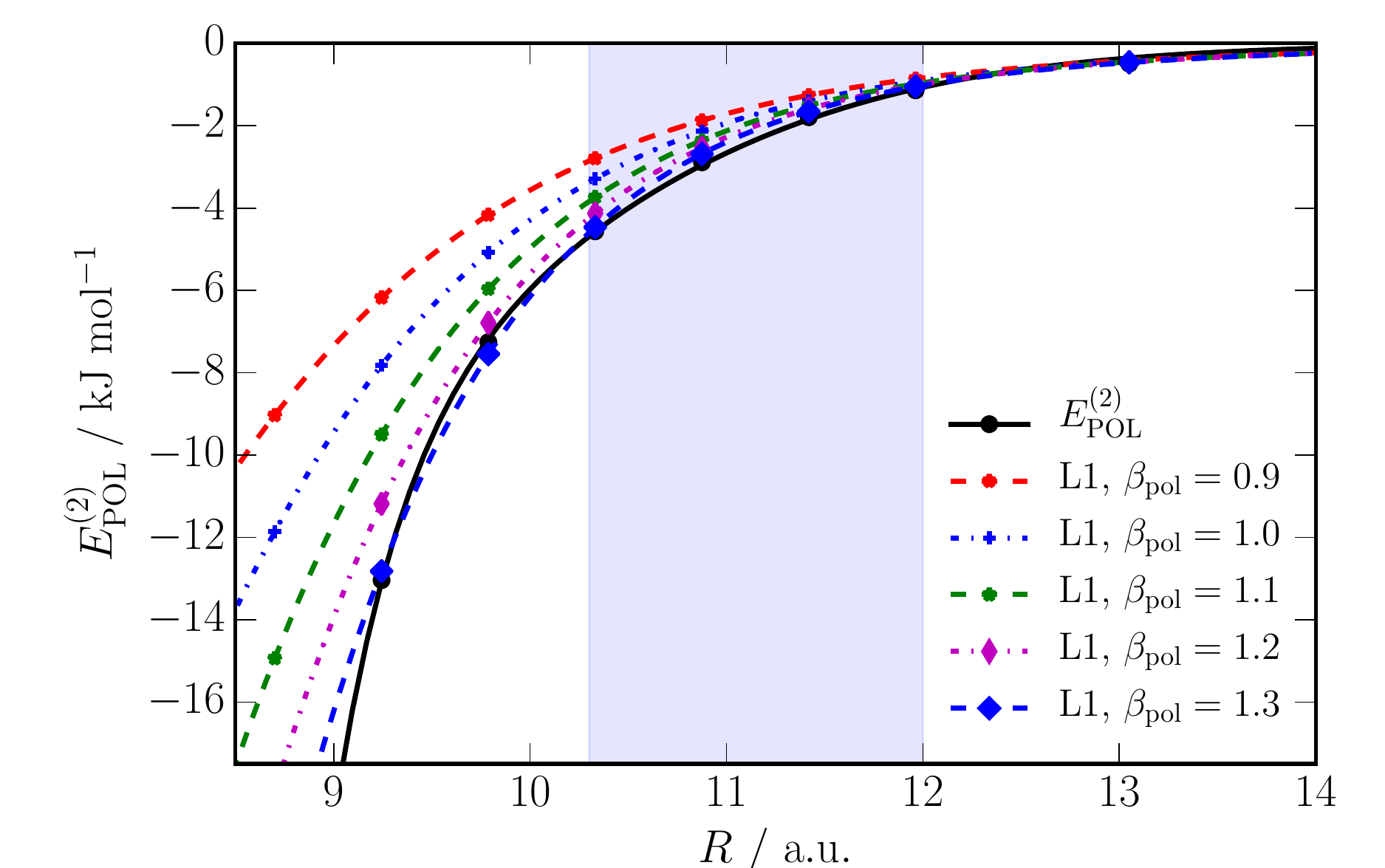}\\
        \includegraphics[scale=0.45]{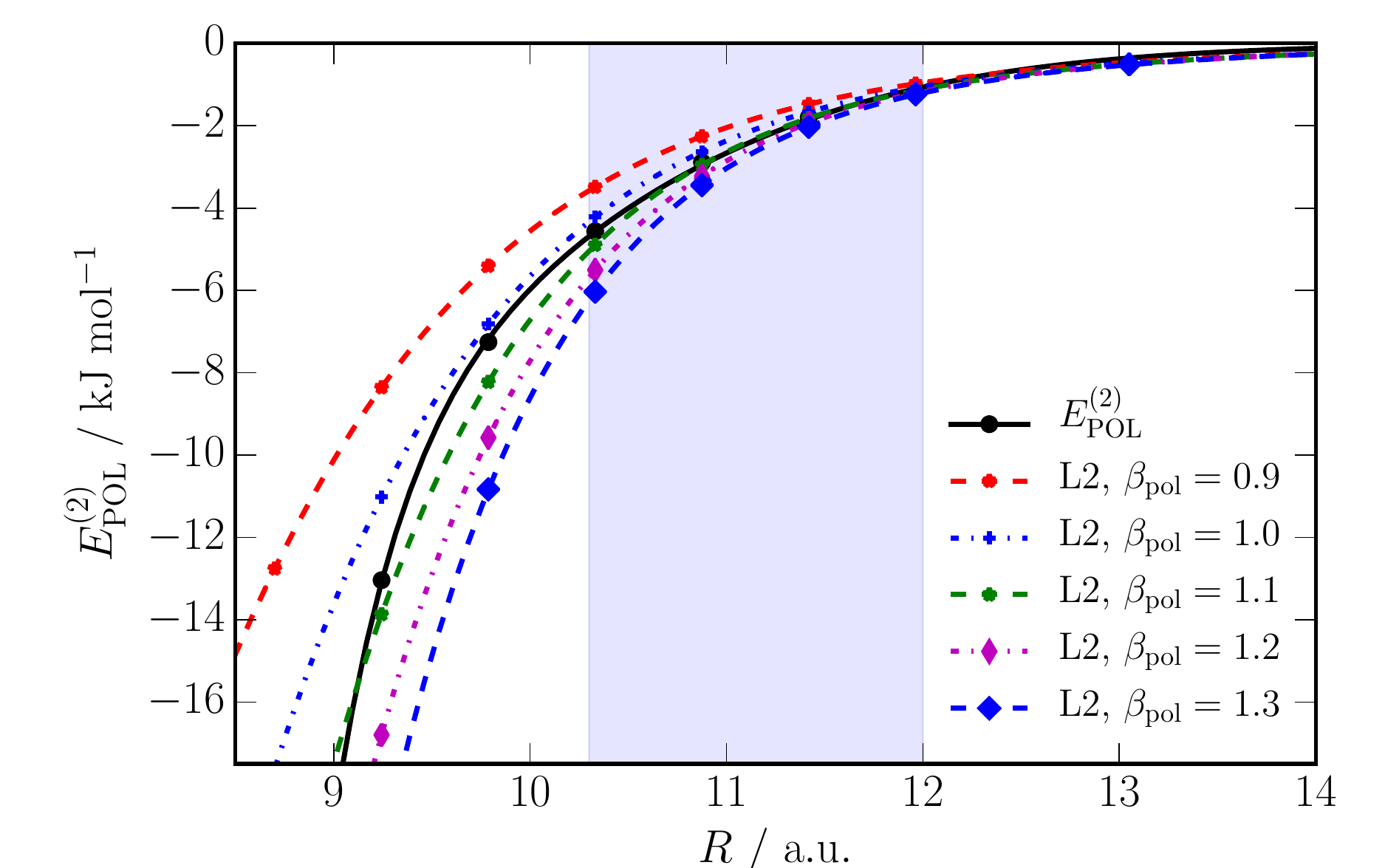}
    \end{center}
\caption{
    Second-order polarization energies vs. centre-of-mass separation
    $R$ for the doubly-hydrogen-bonded pyridine dimer, obtained from
    regularized SAPT(DFT) and from distributed polarization models.
    Model polarization energies are reported with local WSM models with a maximum rank of 
    1 (L1, top) and 2 (L2, bottom). Models are shown with a range of damping 
    parameters using damping functions described in the text.
    The basin of the minimum along the radial direction is indicated by the
    light blue shaded region.
}
\label{fig:pyr2-pol-Hb1}
\end{figure}

\begin{figure}
    \begin{center}
        \includegraphics[scale=0.45]{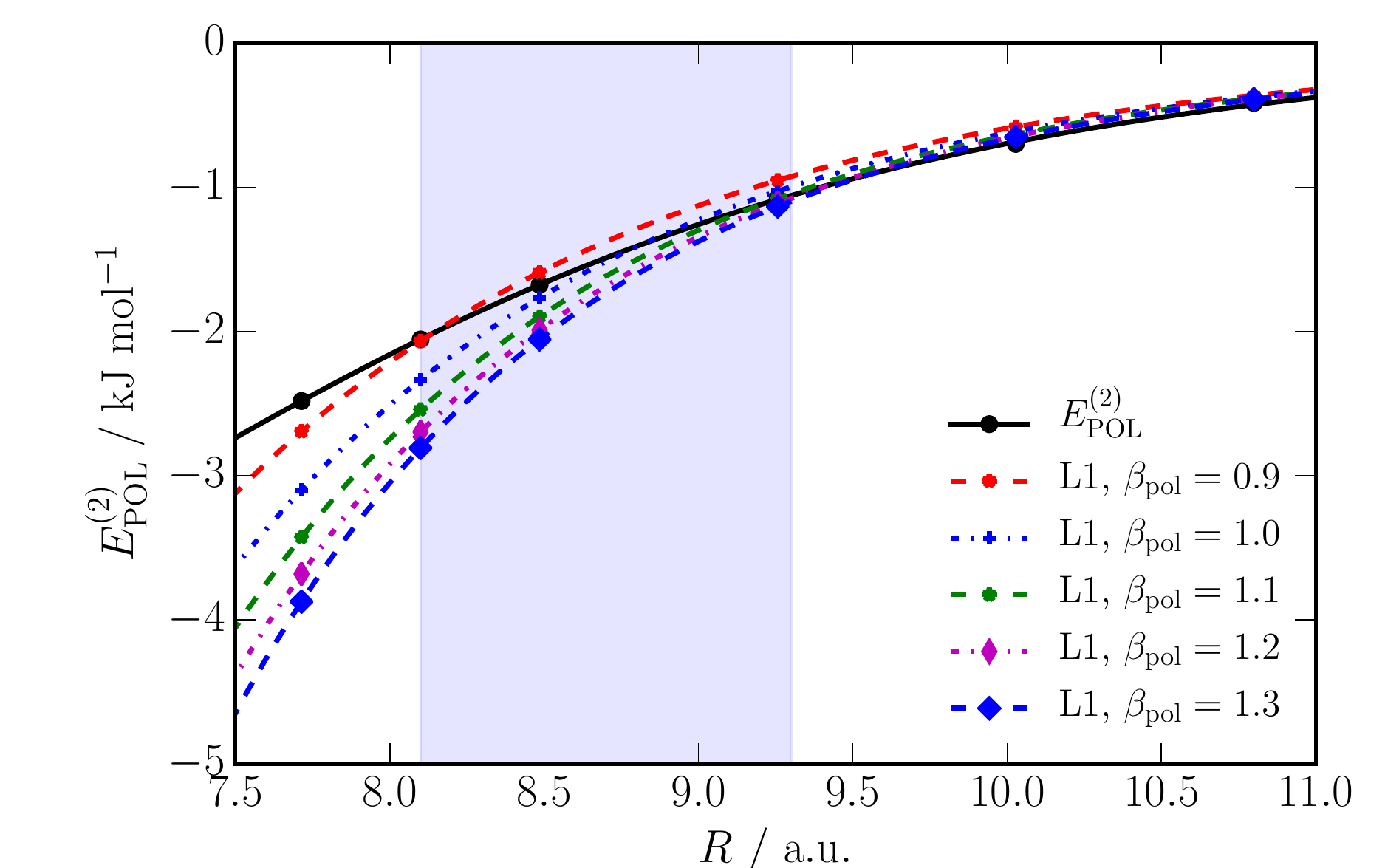}\\
        \includegraphics[scale=0.45]{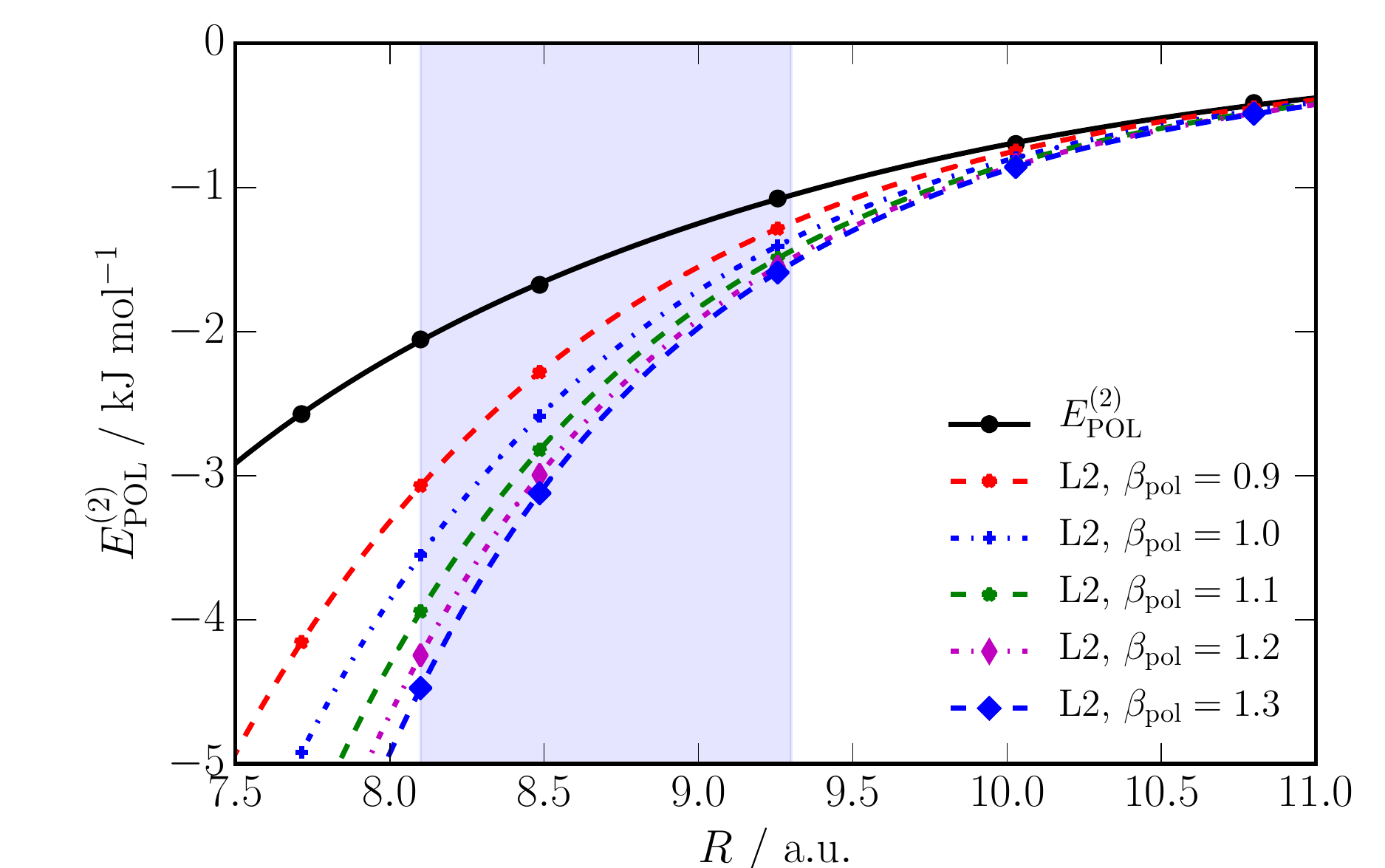}
    \end{center}
\caption{
    Second-order polarization energies vs. centre-of-mass separation
    $R$ for the T-shaped pyridine dimer with N pointing 
    to the centre of the ring.
    See the caption of Figure~\ref{fig:pyr2-pol-Hb1} for a description.
}
\label{fig:pyr2-pol-N-ring}
\end{figure}

In principle the above procedure gives us a straightforward way to define the
damping: once the form of the damping function is chosen (we use
Tang--Toennies damping in this work) all we need to do is 
determine the damping parameters needed by fitting to \EPOL{2} energies calculated
for a suitable set of dimer orientations.
Since the many-body polarization energy is built up from terms involving 
pairs of sites, we should expect that the damping parameters
depend on the pairs of interacting sites, and potentially on their
relative orientations. Indeed, one of us has shown \cite{Misquitta13a} that
for small dimers the damping parameters do depend quite strongly on the
site types involved. 
A single-parameter damping model that depends only on the types of interacting
molecules may be constructed, but such a model is a compromise, and must usually
be determined by fitting to data biased towards the
important dimer configurations only \cite{Misquitta13a}.
The advantage of this approach is that the model is simpler and very few
evaluations of \EPOL{2} are needed to determine the damping parameter, but the
disadvantage is that the model is almost certainly biased towards a few dimer orientations,
and additionally, these important orientations need to be known before the final potential 
is constructed. The last requirement---that we need to have knowledge of the potential---is
not as serious as it may seem, as the choice of damping has no effect on the 
\emph{two-body} interaction potential: this choice affects the many-body polarization
energy only. So it is possible to make an informed guess for the damping parameter,
determine the parameters of the intermolecular potential, and subsequently
re-assess this choice by examining the performance of the polarization model
at the important dimer configurations, and, if necessary, alter the model and re-fit.

The initial choice for the damping parameter in pyridine was obtained using two dimer
orientations: the doubly hydrogen-bonded $C_{2h}$ dimer, and a T-shaped dimer with
the nitrogen of one molecule pointing to the ring of the other. 
These were chosen so as to sample both H$\cdots$N and N$\cdots$C interactions,
though in retrospect the latter proved to be unimportant.
In Figure~\ref{fig:pyr2-pol-Hb1} we display the second-order polarization energies
calculated using various single-parameter damping models for the $C_{2h}$ structure. 
Energies for only two of the three polarization models are shown, as the isotropic
rank 1 (L1(iso)) model is nearly identical in behaviour to the L1 model.
The optimum damping parameter for the L1 model lies between 1.2 and
1.3 a.u., while for the L2 model a stronger damping between 1.0 and 1.1 a.u.\ is needed. 
To an extent, the deficiencies of the L1 model are compensated by using a weaker
damping. 

The single-parameter damping approach has a serious limitation. 
In Figure~\ref{fig:pyr2-pol-N-ring} we display similar data for the T-shaped dimer orientation
with the N of one molecule pointing to the centre of the ring of the other.
Here we see that the polarization models need to be considerably more heavily 
damped with a damping coefficient of 0.9 a.u.\ for the L1 (and L1(iso)) model 
and one less than 0.9 a.u.\ for the L2 model. 
It is possible that we observe this large variation in the damping because
of the strong anisotropy of the molecule, and also because a single damping 
coefficient is not enough. Perhaps we need to use separate damping parameters
for each pair of atoms \cite{Misquitta13a}, or even to make the damping parameters
orientation-dependent.
As a compromise, we have chosen to use the simpler L1 model with a damping 
coefficient of $\betapol = 1.25$ a.u. This model seems capable of describing 
the polarization in both orientations presented here.

This approach to choosing the damping parameter remains the most problematic
part of our approach to potential development. The choice of damping parameters
may seem somewhat arbitrary and biased to the choice of dimer configurations used to determine
the damping, but this is probably too pessimistic a view for the following reasons:
\begin{itemize}
    \item The choice of damping does not affect the two-body interaction energy
        as the error in the induction energy will
        be absorbed in the short-range part of the potential. 
        The damping does however alter the many-body polarization energy.
    \item We should regard this as an iterative process: the damping model
        will normally be assessed and possibly changed once we have a better
        understanding of the full PES. Indeed this was done in the present
        work. We will re-visit this issue in \S\ref{B-sec:pol-damping-revisited} 
        in \paperB.
\end{itemize}

%

\subsection{Dispersion models}
\label{sec:dispersion}

In \CamCASP, we normally calculate atom--atom dispersion coefficients
using polarizabilities computed at imaginary frequency  and localised
using the WSM localization scheme.
The procedure involves integrals over imaginary
frequency\cite{timf96}, and because the imaginary-frequency
polarizability is a very well-behaved function of the imaginary
frequency the integrals can be carried out accurately and efficiently
using Gauss-Legendre quadrature\cite{MisquittaS08b}. Since the
dispersion coefficients are derived from the WSM polarizability model,
it is possible to choose the dispersion model to suit the problem, for
example by limiting the polarizabilities to isotropic dipole--dipole,
leading to an isotropic $C_6R^{-6}$ model, or by including all
polarizabilities up to quadrupole--quadrupole, which yields a model
including anisotropic dispersion terms up to $R^{-10}$. (This latter
procedure omits some $R^{-10}$ terms arising from dipole--octopole
polarizabilities, but they could be included too if desired.)
Within the constraints of the model, the WSM polarizabilities, and
hence the WSM dispersion models will be optimized to be the best
in an unbiased sense. Within these constraints, intramolecular through-space 
polarization effects are fully or partially accounted for in the WSM models.


The dispersion energy of pyridine with a neon atom probe placed on the vdW$\times2$
surface of pyridine is mapped in the left-hand picture of Figure~\ref{fig:maps}(b). 
In the remaining three maps in this figure we show the error in the dispersion 
energy for three local dispersion models: the \Cn{6} model includes anisotropic 
\Cn{6} terms on all atoms; the \Cn{8} model additionally includes \Cn{7} and \Cn{8} 
contributions between the heavy atoms; and the $C_{6,{\rm iso}}$ model includes
only isotropic \Cn{6} terms. The \Cn{10} and \Cn{12} models are not shown as they
exhibit errors close to zero on the scale shown. 
It should be clear that to achieve a high accuracy we need to include higher-rank
dispersion effects --- the dispersion anisotropy is not apparently important in this
system, though we may expect it to be important in larger systems. 
Also, the errors made by both the \Cn{6} models are fairly uniform, and so the lack of
higher-order terms in these models may be compensated for by scaling the \Cn{6}
coefficients. Indeed, we have demonstrated this in a previous publication \cite{MisquittaS08b}
and will address this below. 

\begin{figure}
    \begin{center}
        \includegraphics[scale=0.45]{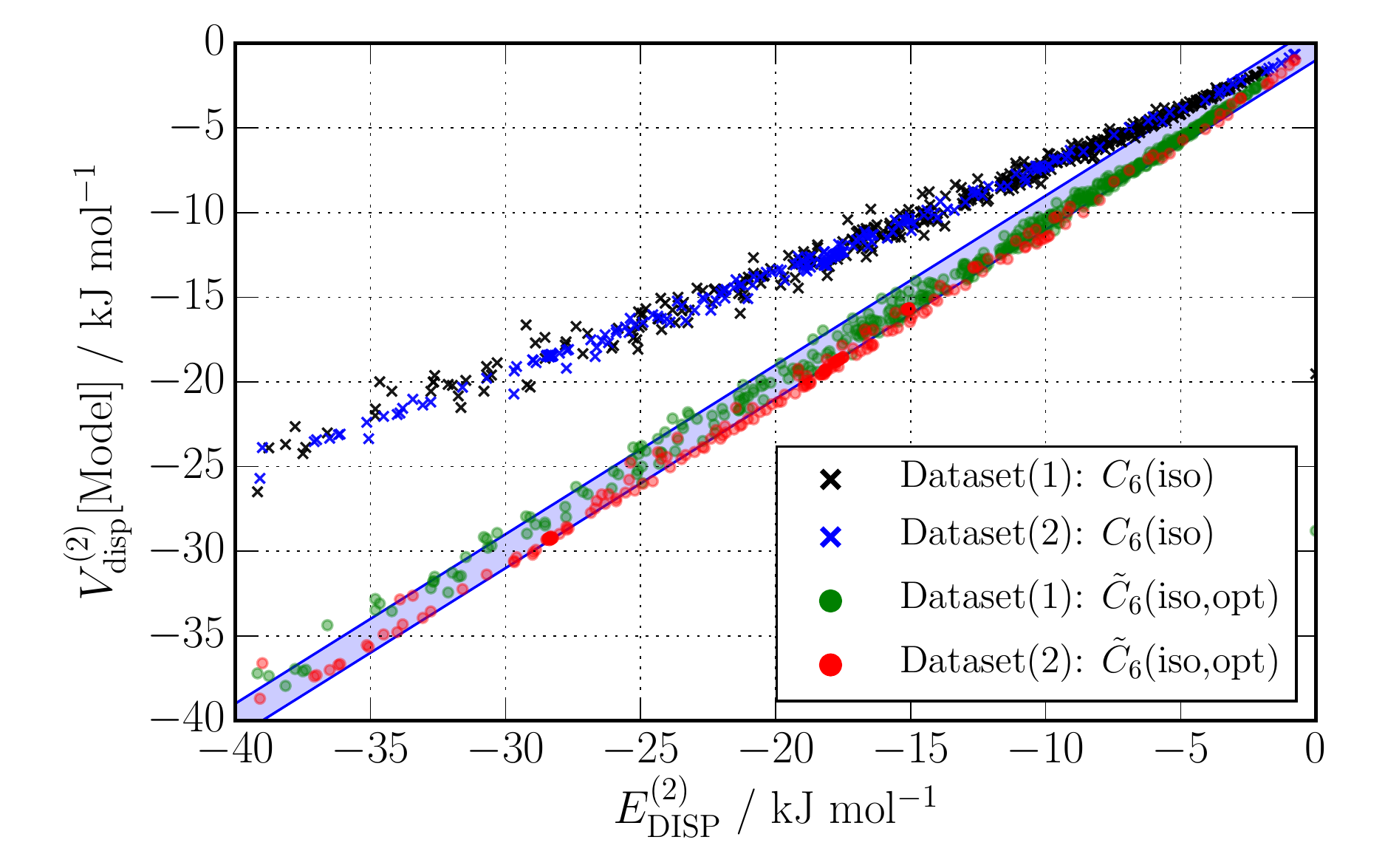}
        \includegraphics[scale=0.45]{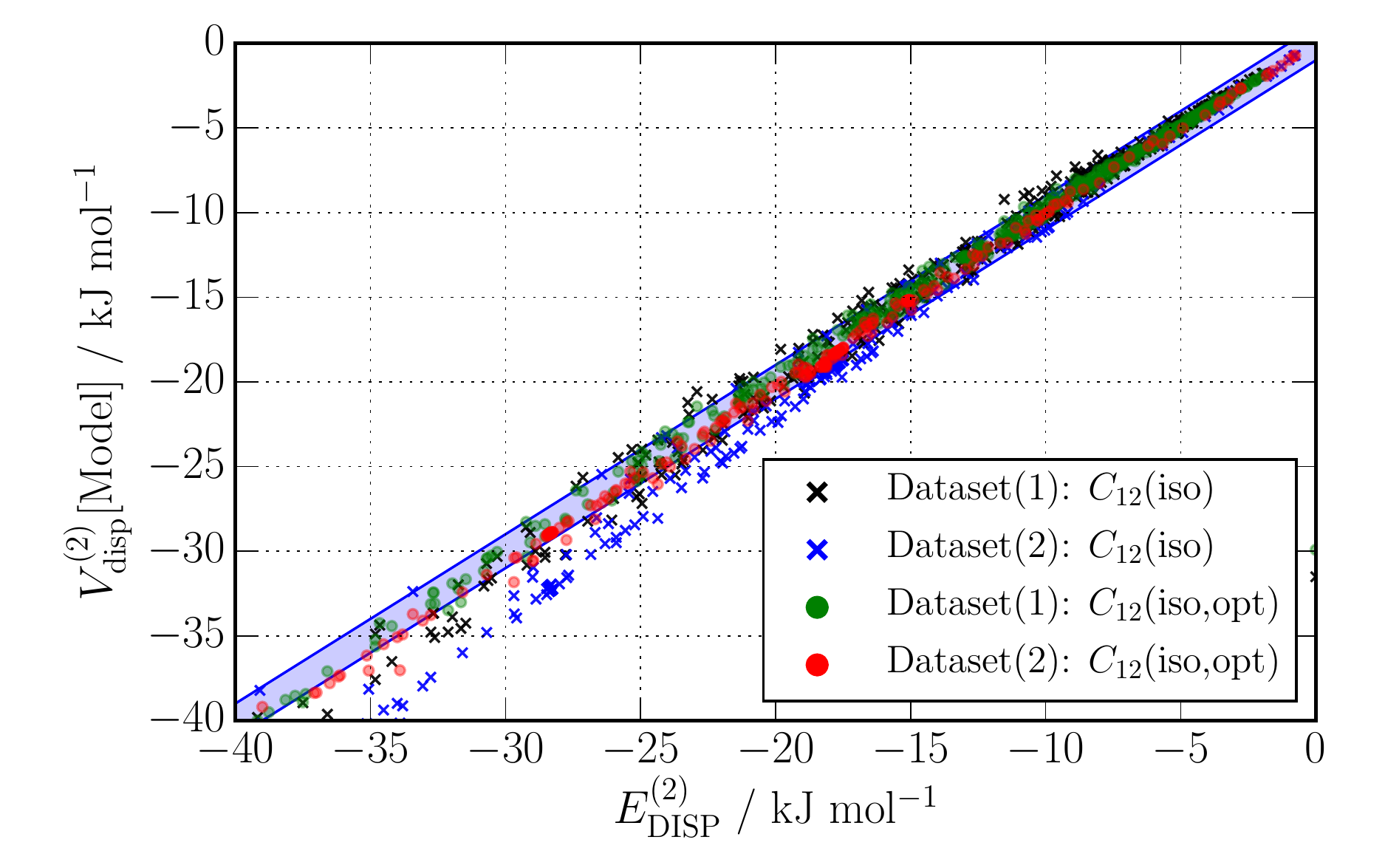}
    \end{center}
    \caption{Scatter plot of pyridine dimer dispersion energies using
        various models, against reference \EDISP{2} energies calculated using
        SAPT(DFT). See text for details. 
        The bar represents the $\pm 1 $ \kJpermol error range.
    \label{fig:disp_scatter}
    }
\end{figure}

The WSM dispersion models described above need to be suitably damped for them
to be applicable in a potential.
We have used the Tang--Toennies \cite{TangT84} damping functions and a single 
damping parameter for all pairs of sites.
The damping model needs to account for two effects:
First, the SAPT(DFT) dispersion energy, \EDISP{2}, includes the effects of penetration
and exchange, which are absent from the $C_n R^{-n}$ expansion. 
Secondly, the dispersion expansion suffers from an unphysical
mathematical divergence as $R \rightarrow 0$. For both reasons the models
have to be damped. Damping using the Tang--Toennies functions
cancels out the mathematical divergence at small $R$ and, with an appropriate
damping parameter, is also able to account for the penetration and 
exchange effects, albeit approximately. 
We have opted for the simplest damping model, in which \betadisp
depends on the interacting molecules only and is given by 
eq.~\eqref{eq:disp-damping}.
With $I_A = I_B =  0.3488$ a.u.\ we get $\betadisp = 1.67 $ a.u. 

Figure~\ref{fig:disp_scatter} (bottom) shows the performance of the
\Cniso{12} isotropic dispersion models for the pyridine dimer,
As can be seen from the Figure, the above damping works reasonably
well for the \Cniso{12} model with (unweighted) \rms errors of $0.86$ \kJpermol for 
dispersion energies from Dataset(1) in the energy range $-40$ to $0$ \kJpermol.
However, for Dataset(2) which includes more strongly bound dimers, the model performs
less well with an \rms error of $2.30$ \kJpermol in the same energy range.
The model dispersion energies are systematically overestimated for the low energy
dimers, with errors as large as $4$ \kJpermol. 
While these errors are just within `chemical accuracy', they are too large for our
purposes.
They may stem from the choice of damping function, the
damping parameter chosen (in particular, our use of a single, atom-pair independent
isotropic damping parameter) and also the WSM dispersion coefficients.
To account for some of these deficiencies, while maintaining the isotropy of the model, 
we have chosen to relax the dispersion coefficients in the \Cniso{12} model.
The relaxation was done using constrained optimisation with harmonic constraints 
in the form given by eq.~\eqref{eq:constrained-opt} used to pin the dispersion
coefficients to the values obtained from the WSM procedure. 
We used tight constraints to prevent the model parameters from taking on
unphysical (negative) values.
The relaxation was done using only the random dimers from Dataset(1), with the low energy 
dimers from Dataset(2) used to assess the quality of the relaxation.
The relaxed model, \CnisoOPT{12}, is a significant improvement, with \rms errors of
$0.41$ \kJpermol on the training set of random dimers and $0.68$ \kJpermol on 
the test set of low energy dimers.

In a similar manner we have created an isotropic \Cn{6} dispersion model for this
system. From Figure~\ref{fig:disp_scatter} (top) we see that the \Cniso{6} model systematically
underestimates the second-order dispersion energy. This is to be expected,
as the higher ranking dispersion contributions are significant
for close dimer separations.
We have previously argued \cite{MisquittaS08a} that rather than use the
\Cniso{6} model directly, we should instead use a \emph{scaled} model in which
all dispersion coefficients are scaled by a constant to match the reference
\EDISP{2} energies. Here we additionally optimise the scaled model in the
manner described above. The resulting model, \tCnisoOPT{6} (here the tilde indicates
that this is a scaled model), exhibits an \rms error of
$0.68$ \kJpermol on the training set and $1.00$ \kJpermol on the test set. 
However, such a scaled model will systematically overestimate the 
long-range contribution to the dispersion energy, and this is a
significant drawback: while the scaled \Cniso{6} model
may be used to model small, gas-phase clusters, it is not suitable for 
the condensed phase because the scaling causes an excessive van der Waals pressure
and the resulting structures are significantly more dense. As one of our aims 
is to use the resulting potentials in the study of the condensed phase, we 
cannot use the scaled model. However, we can simplify the \Cniso{12}
model by dropping the $R^{-12}$ terms, which 
contribute an insignificant amount to the dispersion energy, so we have
used a \CnisoOPT{10} model in the potentials for pyridine.

\section{Short-range energy models}
\label{sec:shortrange}

The short-range part of the potential comprises several effects. All
of the long-range terms are modified at short range, as mentioned
above. The multipole expansion on which the long-range expressions are 
based converges more slowly or not at all at short distances, and is
incorrect when the charge densities overlap, even if it does converge.
Damping can be used to correct the dispersion and polarization
terms at short range, but
in addition there are corrections arising from electron exchange,
electrostatic penetration, and charge tunneling, or charge transfer,
between the molecules. 

The dominant short-range term is the exchange-repulsion: the
wavefunction for two overlapping molecules cannot be treated as a
simple product of isolated-molecule wavefunctions, but has to be
antisymmetrized with respect to electron exchanges between the
molecules. This modifies the electron distribution and results in a
repulsive energy. It is straightforward to calculate the
exchange-repulsion energy \abinitio, but it has to be fitted by a
suitable functional form for use in an analytic potential.

The electrostatic interaction is also modified by the effects of
overlap. If a distributed multipole expansion is used, it will still
converge at moderate overlap, but it does not converge to the non-expanded
energy, \Eelst. The difference between \Eelst and the converged multipole
energy \EelstMP is the electrostatic penetration energy, \Eelstpen.
We have previously shown \cite{MisquittaSF14} that \Eelstpen is 
approximately proportional to the first-order exchange energy, so
the two terms can, in principle, be modelled together. 
Alternatively a separate model for \Eelstpen can be developed, possibly
based on suitable damping functions \cite{Stone11}, but we have not 
explored this possibility.

The contribution to the interaction energy from charge transfer ---
or, more appropriately, the intermolecular charge delocalisation
energy ---
appears at second and higher orders in the perturbation expansion.
Previously one of us has shown that this energy can be interpreted as an
energy of stabilization due to electron tunneling \cite{Misquitta13a},
so we may expect the charge transfer energy to decay
exponentially with separation. In principle, the charge transfer
energy should be modelled as a separate exponentially decaying term, but
as we shall see, it is approximately proportional to the first-order
exchange energy and may therefore
be modelled together with \Eexch.

Finally we will use the short-range potential to account for any residual
differences between the multipole expansions and the reference SAPT(DFT) energies.
The full form of the short-range energy, \Esr{1-\infty}, is shown in 
eq.~\eqref{eq:Esr} where we have also implicitly defined the first-order
short-range energy, \Esr{1}, and the contributions from second to 
infinite order, \Esr{2-\infty}.

\subsection{Fitting the short-range potential}
\label{sec:sr-fitting}

The short-range part of the potential has often been represented by
empirical $R^{-12}$ Lennard-Jones atom--atom terms, but for accurate
potentials a Born--Mayer (exponential) atom--atom form is usually
preferred (eq.~\eqref{eq:Vsr}), and it is essential in most cases to allow it to be
anisotropic, since the non-spherical nature of bonded atoms can have a
profound effect on the way that they pack together.
Unfortunately, the
parameters of the various atom--atom terms are strongly correlated, and
this makes the already difficult non-linear fitting problem
even more troublesome. A direct fit is generally not possible: it is hard
to converge and tends to wander off into unphysical parameter space.
Parameters can be forced to stay within reasonable limits, but this introduces
an element of arbitrariness in the procedure. 

It has however been found empirically that there is a close
proportionality between the overlap of the electron densities on two
atoms and the exchange--repulsion energy between them. This
observation has been used to construct repulsion potentials
directly from the density overlap, with varying degrees of
success\cite{SoderhjelmKR06}. A 
better solution, which we adopt here, is to use the density overlap
only to guide the parameters in a fitted potential function to a
physically meaningful region of parameter space. Once an initial guess to the parameters
has been obtained, the fit can be improved using constrained optimisation.
Further, we will achieve the final fits to \Esr{1-\infty} in stages, first
by fitting to only \Esr{1}, and then by constrained relaxation to include the
higher-order contributions from \Esr{2-\infty}.

\subsection{The density-overlap model}
\label{sec:dens-overlap-model}

It is useful at this point to review the theoretical basis for the
density-overlap model.
In the mid-1970's Kita, Noda \& Inouye \cite{KitaNI76}, and later, in the early 1980's 
Kim, Kim \& Lee\cite{KimKL81} proposed that the
intermolecular repulsion energy of rare gas atoms could be modelled as 
\begin{align}
    \Eexch(\RR) \approx K \bigl(S_{\rho}(\RR)\bigr)^{\gamma},
        \label{eq:overlapmodel}
\end{align}
where $K$ and $\gamma$ are constants and the overlap $S_{\rho}$ of the
two interacting densities $\rho^{A}$ and $\rho^{B}$ separated by generalised
vector $\RR$ is defined as
\begin{align}
  S_{\rho}(\RR) = \int \rho^{A}(\rr) \rho^{B}(\rr) d\rr.
   \label{eq:overlap}
\end{align}
Kita \etal had $\gamma=1$ and did not consider the possibility of varying this power,
but Kim \etal observed that the constant $\gamma$ was close to, but less than, unity.
This model was subsequently used by a number of groups and was successfully
applied to study the interactions of polyatomic molecules, and has been
investigated \cite{MitchellP00,SoderhjelmKR06} together with many other variants.
Curiously, to the best of our knowledge, no one seems to have realised the 
reason for the success of this model, nor why the constant $\gamma$ is always
less than one. Before going on to the numerical details of
this model we will discuss both these issues as we will be led to a better
understanding of the model and the exchange-repulsion energies.

First of all we should realise that although the exchange-repulsion and
penetration energies are the short-range parts of the interaction energy, 
these energies result from the overlap of the \emph{density tails} of the
interacting densities. That is, we must consider the asymptotic form of the
interacting densities for an atomic system \cite{PatilT-AsymptoticMethods}:
\begin{align}
  \rho(r) = C r^{2 \beta} e^{-2 \alpha r},
  \label{eq:rhoasymp}
\end{align}
where, with $I$ as the vertical ionization energy, and $Z$ the atomic number,
we have $\alpha = \sqrt{2 I}$ and $\beta = -1 + Q/\alpha$, where for an
atom with nuclear charge $+Z$ and electronic charge $-N$, $Q=Z-N+1$.
Both $I$ and $r$ here are in atomic units.
In principle, the asymptotic form of the density overlap integral can
be obtained by using this density in eq.~\eqref{eq:overlap}, but the
exact integral is not important. Instead we can use the result of
Nyeland \& Toennies \cite{NyelandT86} who evaluated
eq.~\eqref{eq:overlap} using only the exponential term in
eq.~\eqref{eq:rhoasymp} to get
\begin{align}
  S_{\rho}(R) = \mathcal{P}(R) e^{-2 \alpha R},
  \label{eq:Sasymp}
\end{align}
where $\mathcal{P}(R)$ is a low-order polynomial in the internuclear separation $R$. 
For identical densities $\mathcal{P}(R) = (4/3) \alpha^2 R^2 + 2 \alpha R + 1$, and 
for the more general case of different densities, the results of Rosen\cite{Rosen31}
may be used to obtain a closed-form expression for $\mathcal{P}(R)$ that is now
not a low order polynomial, but also includes exponential terms.
Since $S_{\rho}$ is not a pure exponential, Nyeland \& Toennies argue that the
exchange-repulsion energy should be proportional to $S_{\rho}(R)/R^2$,
but this assumes that the exchange-repulsion itself is a pure
exponential, which is not the case.

The asymptotic form of the exchange-repulsion energy has been worked out 
by Smirnov \& Chibisov \cite{SmirnovC65} using the surface-integral approach
and later, with a corrected proof, by Andreev \cite{Andreev73}.
Their result is 
\begin{align}
  \Eexch = K R^{(7/2\alpha)-1} e^{-2\alpha R}
\label{eq:asymp_er}
\end{align}
where $K$ is an angular momentum-dependent constant \cite{KleinekathoferTTY95}.
We observe that:
\begin{itemize}
    \item The exchange-repulsion energy is not a pure exponential, as is often
        assumed, but is better represented as an exponential times a function of $R$.
        This has been empirically verified by \citet{ZemkeS99} using
        spectroscopic data for alkali diatomic molecules. Also, accurate analytic
        potentials for small van der Waals complexes have tended to use functional
        forms that include a pre-exponential polynomial term
        \cite{KoronaWBJS97,MisquittaBS00,BukowskiSC99}.  The prefactor function in
        eq.~\eqref{eq:asymp_er} is not a polynomial, but it is close to linear in $R$
        for relevant values of $\alpha$ and $R$.
    \item The exchange-repulsion energy has an asymptotic form that is very similar
        to that of the density overlap, eq.~\eqref{eq:Sasymp}, but the prefactor
        is different. Consequently we should not expect a direct proportionality between the
        two, and a better form of the density-overlap model might use
        \begin{align}
          \Eexch(\RR) \approx \mathcal{K}(\RR) S_{\rho}(\RR),
        \end{align}
        where $\mathcal{K}(\RR)$ is a low-order polynomial in $\RR$.
    \item The exponents in the asymptotic forms of the density overlap and the
        exchange--repulsion will be the same only if the wavefunctions used to
        evaluate them are the same. In general this will not be the case. While the 
        exchange--repulsion could be evaluated with electron 
        correlation effects included, the density-overlap integrals are more
        typically evaluated using Hartree--Fock densities.
        Therefore, the $\alpha$ in the exponent of eq.~\eqref{eq:Sasymp} must be replaced
        by $\alpha_{\text{HF}} = (-2 \epsilon_{\text{HOMO}})^{1/2}$, where 
        $\epsilon_{\text{HOMO}}$ is the energy of the highest occupied molecular orbital
        from Hartree--Fock theory.
        In this case, there will be a better agreement between the exchange--repulsion
        energy and the density overlap if the exponents are made the same by raising
        the latter by the power $\gamma = (-I/\epsilon_{\text{HOMO}})^{1/2}$ as
        is done in eq.~\eqref{eq:overlapmodel}. Now in Hartree--Fock theory 
        $|\epsilon_{\text{HOMO}}| \gt I$, so $\gamma$ is always less than unity, and for
        the helium, neon and argon dimers we obtain values between 0.99 and 0.97 
        in reasonable agreement with the empirical results of Kim \etal.
\end{itemize}
We will now use these observations to construct models for the short-range
energies. 

Electron charge densities obtained from density functional theory are exact,
in principle. In practice, because of the now well understood
self-interaction problem with standard local and semi-local
exchange-correlation functionals, they tend to be too diffuse. This
can be corrected by applying a suitable asymptotic correction to the
exchange-correlation potential \cite{TozerH98,GruningGGB01}. It is now usual to
apply this correction in any SAPT(DFT) calculation; without it, even
energies that depend on the unperturbed monomer densities, like the 
electrostatic energy, can be significantly in error. With the asymptotic
correction, the asymptotic form of the density given by eq.~\eqref{eq:rhoasymp}
is enforced, and consequently $\gamma=1$ in eq.~\eqref{eq:overlapmodel}.

This has important consequences for multi-atom systems where we use the 
overlap model to partition \Eexch into contributions from pairs of atoms.
This idea goes back to the work of Mitchell \& Price \cite{MitchellP00}
and begins with a partitioning of the densities into spatially localised
contributions that will usually be centered on the atomic locations.
If we can write
\begin{align}
    \rho^{A}(\rr) = \sum_a \rho^{A}_{a}(\rr),
      \label{eq:density-partitioning}
\end{align}
where $\rho^{A}_{a}$ is the partitioned density centered on (atomic) site $a$,
and likewise for $\rho^{B}$, then from eqs.~\eqref{eq:overlapmodel} and \ref{eq:overlap}
we get
\begin{align}
    \Eexch(\RR) &\approx \sum_{ab} K  \int \rho^{A}_{a}(\rr) \rho^{B}_{b}(\rr) d\rr \nonumber \\
                &\approx \sum_{ab} K S_{\rho}^{ab}(\RR),
\end{align}
where $S_{\rho}^{ab}$ is the site--site density overlap. This expression may be
generalised by introducing a site-pair dependence on $K$ as follows:
\begin{align}
    \Eexch(\RR) &\approx \sum_{ab} K_{ab} S_{\rho}^{ab}(\RR)
            = \sum_{ab} \Eexch[ab](\RR),
      \label{eq:dist-overlapmodel}
\end{align}
where $\Eexch[ab]$ is the first-order exchange contribution assigned to 
site-pair $(ab)$. This is the distributed density overlap model. 
This is essentially the result obtained by Mitchell \& Price but in their case,
because of their use of electronic densities from Hartree--Fock theory, 
they had $\gamma \lt 1$ and so obtained an expression for the partitioning that is
necessarily approximate. 

There are a few important issues about the overlap model given in
eq.~\eqref{eq:dist-overlapmodel}:
\begin{itemize}
    \item The model was originally formulated for the first-order exchange
        repulsion only, but, as the other short-range energy contributions
        are also roughly proportional to \Eexch, we may use the density-overlap
        model more generally for all of the short-range energy,
        \Esr{1-\infty}. Henceforth we will use the model in this general sense,
        that is, to model the short-range energy, \ESR, however we may choose to 
        define it.
    \item The model allows us to partition the short-range energy into terms 
        associated with pairs of sites. With this partitioning, we may fit
        an analytical potential to individual site pairs rather than fit
        the sum of exponential terms given in eq.~\eqref{eq:Vsr}.
        The fit to each individual term $\Vsr{}{ab}$ (eq.~\eqref{eq:Vtot_Vsr})
        is numerically better defined and may be achieved with relative ease.
    \item This is an approximation: Since the density overlap model
        cannot exactly model the short-range energy, we have 
        $\ESR(\RR) \ne \sum_{a,b} \ESR[ab](\RR)$.
        That is, there is a residual error that originates from the
        original ansatz given in eq.~\eqref{eq:overlapmodel}.
    \item Although the residual error is small compared with \ESR,
        it needs to be accounted for to achieve an accurate fit,
        particularly as the error may be a non-negligible fraction of the
        total interaction energy, which is generally much smaller in
        magnitude than \ESR.
        This may be achieved by constrained relaxation of the final 
        short-range potential $\VSR{} = \sum_{ab} \Vsr{}{ab}$.
\end{itemize}

\section{ISA-based distributed density overlap}
\label{sec:ISA-density-overlap}

Formally, the distributed density overlap integrals, $S_{\rho}^{ab}(\RR)$,
defined through eqns.~\eqref{eq:density-partitioning} and \eqref{eq:dist-overlapmodel},
are particularly straightforward to evaluate using the BS-ISA algorithm 
\cite{MisquittaSF14} as this algorithm provides basis-space expansions for the
atomic densities $\rho^{A}_{a}(\rr)$. However, basis-set limitations
mean that while the BS-ISA algorithm 
results in fairly well-defined atomic shape-functions, the atomic
densities are not well described in the region of the 
atomic density tails, where the density can even attain small negative
values. This not only leads to distributed density overlap integrals that can be
negative, but also results in a relatively poor correlation between the 
first-order exchange energies and the density overlap integrals.
This problem may be alleviated using better basis sets for the atomic 
expansions, but we have not as yet explored this option.

An alternative is to evaluate $S_{\rho}^{ab}(\RR)$ using the atomic densities
defined as
\begin{align}
    \rho^{A}_{a}(\rr) &= \rho^{A}(\rr) \times \frac{\wT{a}(\rr)}{\sum_{a'} \wT{a'}(\rr)},
    \label{eq:atomic-density}
\end{align}
where $\wT{a}$ is the tail-corrected shape-function for site $a$ as defined
in Ref.~\citenum{MisquittaSF14} as a piece-wise function:
\begin{align}
    \wT{a}(\rr) = 
        \begin{cases}
            \w{a}(\rr)  & \text{if} \ |\rr| \leq r^a_0 \\
            \wL{a}(\rr) & \text{otherwise},
        \end{cases}
\end{align}
where $\w{a}(\rr)$ is the atomic shape-function that is the
spherical average of atomic density $\rho^{A}_{a}(\rr)$, and
the long-range form of the shape-function is defined as
$\wL{a}(\rr) = A_{a} \exp{(-\alpha_{a} |\rr - \RR_{a}|)}$, where
$r^a_0$ is a cutoff distance, and the constants in \wL{a} are defined
to enforce continuity and charge-conservation. \cite{MisquittaSF14}
The shape-functions may be thought of as pro-atomic densities that
encode the ionic state of the atom in its molecular environment. This
ionic state is not fixed and is instead determined self-consistently
through the ISA iterations \cite{LillestolenW09}.
While the atomic shape-functions are spherically symmetrical, the atomic
densities are not.
Now, the distributed density overlap integral is defined as 
\begin{align}
    S_{\rho}^{ab}(\RR) = \int \left( \rho^{A}(\rr) \frac{\wT{a}(\rr)}{\sum_{a'} \wT{a'}(\rr)} \right)
                              \left( \rho^{B}(\rr) \frac{\wT{b}(\rr)}{\sum_{b'} \wT{b'}(\rr)} \right)
                              d\rr.
                              \label{eq:dist-densovr-corr}
\end{align}
Due to the piece-wise nature of \wT{a}, this integral must be evaluated
numerically using a suitable atom-centered integration grid. 
Using techniques described by us earlier \cite{MisquittaSF14}, we evaluate the
terms in eq.~\eqref{eq:dist-densovr-corr} in $\order{N^0}$ computational effort.
This is done by defining local neighbourhoods, \neighbourhood{a} and \neighbourhood{b},
for sites $a$ and $b$. These neighbourhoods are based on the dimer configuration, so 
\neighbourhood{a} may include sites that belong to monomer B, and vice versa for
\neighbourhood{b}.
The neighbourhoods are usually defined using an overlap criterion that naturally takes 
the basis set used into account with basis sets containing more diffuse functions
resulting in larger neighbourhoods.
The integration grid, and various terms in the integral $S_{\rho}^{ab}$ are then
evaluated using sites in the intersection set $\neighbourhood{a} \cap \neighbourhood{b}$.
This intersection set may be null for monomers that are sufficiently far apart. 
In this manner the density overlap integrals are calculated with linear effort.

\section{Summary}
\label{sec:conclusions}

This completes the overview of the method that we have applied to the
pyridine dimer in the following paper.
To summarize, we have described a robust and easily implemented algorithm for 
developing accurate intermolecular potentials in which most of the potential
parameters are derived from the charge density and density response functions.
Significantly, the remaining, short-range parameters are robustly determined
by associating these with specific atom pairs using a distributed density-overlap
model based on a basis-space implementation of the iterative stockholder atoms
(ISA) algorithm. 

We have developed multipole expanded models for the electrostatic,
polarization and dispersion interactions for the pyridine dimer. 
The electrostatic model is based on a distributed multipole analysis (DMA)
that uses a density partitioning method based on the basis-space version of the
iterated stockholder atoms algorithm (BS-ISA) \cite{MisquittaSF14}.
These ISA-DMA multipoles are demonstrated to converge more rapidly with rank that
the more commonly used distributed multipoles from Stone's algorithm \cite{Stone05b},
and additionally, the ISA-DMA expansion is shown to demonstrate a systematic decrease in
accuracy when truncated to lower ranks. In \paperB we will probe these properties of the
ISA-DMA expansion on the total interaction energy models for the pyridine dimer.

We have used regularised SAPT(DFT) \cite{Misquitta13a} to develop a polarization 
models for the pyridine dimer. The L1 model includes anisotropic distributed 
polarizabilities and the L2 model additionally includes rank 2 terms on the 
heavy atoms. Both models have been damped to recover the second-order
true polarization energy defined as the regularised second-order induction 
energy \cite{Misquitta13a}. 
In both cases the distributed polarizabilities have been obtained using the 
Williams--Stone--Misquitta (WSM) algorithm \cite{MisquittaS06,MisquittaS08a,MisquittaSP08,MisquittaS08b}.
This algorithm has also been used to develop distributed dispersion energy models 
for the pyridine dimer. We have tuned these models --- one with \Cniso{6} terms only
and the other including terms to \Cniso{12} --- to SAPT(DFT) total dispersion 
energies. 

We have in addition provided an argument based on the asymptotic forms
of the first-order exchange-repulsion energy and the density-overlap which 
provides a theoretical explanation for the success of the density-overlap model.
Additionally, we have demonstrated that the power $\gamma$ used in the 
density overlap model should be identically $1$ if asymptotically
correct densities are used. 
Setting $\gamma=1$ allows the density-overlap model to be distributed
so as to partition the short-range energy into terms associated with pairs
of sites. This distribution has been used before by other groups, but here
we base it on a firm theoretical foundation.
We argue that while the exponential
terms in the first-order exchange energy and the density-overlap agree,
the polynomial pre-factors are different, so that a better model
may be achieved by allowing the model to contain a distance-dependent
pre-factor. 

Finally, we have provided an algorithm for a distributed density-overlap model for
the short-range (repulsion) energy that used the BS-ISA density-partitioning
scheme rather than the density-fitting scheme we have advocated in previous 
papers \cite{StoneM07,MisquittaWSP08}. In the \paperB, we will demonstrate how, 
with this algorithm, in particular the ISA approach to atoms-in-a-molecule,
a set of accurate, many-body potentials for the pyridine dimer can be derived using a 
relatively small number of dimer energies calculated using SAPT(DFT).
Importantly, we will demonstrate how with this approach we resolve the difficulties
hitherto encountered in determining the short-range parameters and the atomic shape
anisotropy terms.

\section{Acknowledgements}
AJM would like to thank Prof Sally Price for initiating this project
and supporting it in its early stages and Dr Richard Wheatley for
useful discussions related to the ISA.
We would like to thank Mary J. Van Vleet for useful comments on the manuscript.
AJM would also like to thank Queen Mary University of London for
computing resources and the Thomas Young Centre for a stimulating
environment.

This work was partially funded by EPSRC grant EP/C539109/1.

\section*{Appendices}
\appendix

\section{Programs}
\label{sec:programs}

Many of the theoretical methods described in this paper are implemented
in programs available for download. Some of these, together with their main
uses in the present work, are:
\begin{itemize}
  \item \CamCASP 5.9 \cite{CamCASP}: 
  Calculation of WSM polarizabilities, the dispersion models, the SAPT(DFT)
  energies, and overlap models.
  \item \ORIENT\ 4.8 \cite{Orient}: Localization of the
  distributed polarizabilities, calculation of dimer energies using the
  electrostatic, polarization and dispersion models, visualization of the
  energy maps, and fitting to obtain the analytic atom--atom potentials.
  \item {\sc Dalton 2.0} \cite{DALTON2}: DFT calculations. 
  A patch \cite{SAPT2008} is needed to enable {\sc Dalton 2.0} 
  to work with \CamCASP.
\end{itemize}

\section{\CamCASP}
\label{sec:CamCASP}

Many of the algorithmic details of the electronic structure methods
implemented in the \CamCASP suite of programs have been described
in previous publications. Rather than provide an exhaustive list, we will 
indicate those algorithms and methods of importance for potential
development, as well as some numerical techniques that are particularly
important for accuracy and computational efficiency.

Some of the capabilities of the \CamCASP suite of programs are as follows:
\begin{itemize}
    \item {\em SAPT(DFT)}: Interaction energies to second-order can be calculated
        using SAPT(DFT) \cite{MisquittaS02,MisquittaJS03,MisquittaS05,MisquittaPJS05b}.
        Infinite-order effects may be approximated using the \deltaHF correction. 
    \item {\em Distributed multipole models}: These may be evaluated using
        both the GDMA algorithms \cite{StoneA85,Stone05b}, or directly 
        from a density-fitting-based partitioning using a variety of constraints
        (see the \CamCASP User's Guide for details), or from the recently 
        implemented ISA algorithm \cite{MisquittaSF14}.
    \item {\em Distributed frequency-dependent polarizabilities}: These may be
        calculated in non-local form using constrained density-fitting-based
        partitioning schemes \cite{MisquittaS06}, which include the SRLO method
        \cite{RobS13a} as a special case. Localised models may be obtained
        using the Williams--Stone--Misquitta (WSM) model 
        \cite{MisquittaS08a,MisquittaSP08}.
    \item {\em Distributed dispersion models}: These may be evaluated directly
        using the non-local frequency-dependent models \cite{MisquittaSSA10}, or from
        localised polarizability models obtained using the WSM procedure 
        \cite{MisquittaS08b}. Models may be isotropic or anisotropic.
    \item {\em Linear-response kernel}: The code is able to evaluate the 
        linear-response kernel using the ALDA, CHF and hybrid, ALDA+CHF, kernels.
        These integrals are evaluated internally.
    \item {\em Interfaces}: \CamCASP can use molecular orbitals calculated from 
        the {\sc Dalton} program (versions from 2006 to 2015 are supported),
        the \NWChem program and \Gamess. 
\end{itemize}
These are the major features of the \CamCASP program, and the code additionally
includes other algorithms that are important for model development. 
These include the ability to calculate distributed density-overlap integrals
and, from these, develop density overlap models for the short-range
intermolecular interaction energy, and interfaces to the \Orient program\cite{Orient}
to aid in visualisation of the interaction energy models and fitting of
intermolecular potentials.

\setlength\bibsep{2pt}
\providecommand{\latin}[1]{#1}
\providecommand*\mcitethebibliography{\thebibliography}
\csname @ifundefined\endcsname{endmcitethebibliography}
  {\let\endmcitethebibliography\endthebibliography}{}

\end{document}